\documentclass[10pt,journal]{IEEEtran}

\ifCLASSINFOpdf

\else

\fi

\usepackage{amsmath}
\usepackage{amsthm}
\newtheorem{Definition}{Definition}
\newtheorem{Theorem}{Theorem}
\newtheorem*{Proof}{Proof}
\newtheorem{Remark}{Remark}
\usepackage{svg}
\usepackage{subfigure}
\usepackage{amssymb}
\usepackage{booktabs}
\begin{document}

\title{Information Dynamics in Evolving Networks Based on the Birth-Death Process: Random Drift and Natural Selection Perspective}
\author{Minyu Feng,~\IEEEmembership{Senior Member,~IEEE,}
        Ziyan~Zeng,~\IEEEmembership{Student Member,~IEEE,}
        Qin Li, 
        Matja\v{z}~Perc,~\IEEEmembership{Member,~IEEE,}
        J\"{u}rgen Kurths
\thanks{ Manuscript received 27 September 2023; revised 10 January 2024; accepted 8 April 2024. M.F., Z.Z., and Q.L. were supported by the National Natural Science Foundation of China (NSFC) (Grant No.62206230) and the Natural Science Foundation of Chongqing (Grant No. CSTB2023NSCQ-MSX0064). M.P. was supported by the Slovenian Research Agency (Grant Nos. P1-0403 and J1-2457). \textit{(Corresponding authors: Minyu Feng and Qin Li)}}
\thanks{Minyu Feng and Ziyan Zeng are with the College of Artificial
Intelligence, Southwest University, Chongqing, China. (email: myfeng@swu.edu.cn)}
\thanks{Matja\v{z}~Perc is with Faculty of Natural Sciences and Mathematics, University of Maribor,
Koro{\v s}ka cesta 160, 2000 Maribor, Slovenia, with Community Healthcare Center Dr. Adolf Drolc Maribor, Vo{\v s}njakova
ulica 2, 2000 Maribor, Slovenia, with Complexity Science Hub Vienna, Josefst{\"a}dterstra{\ss}e 39, 1080
Vienna, Austria, with Department of Physics, Kyung Hee University, 26 Kyungheedae-ro,
Dongdaemun-gu, Seoul 02447, Republic of Korea.}
\thanks{Qin Li is with the School of Public Policy and Administration, Chongqing University, Chongqing, China. (e-mail: qinli1022@gmail.com)}
\thanks{J\"{u}rgen Kurths is with the Potsdam Institute for Climate Impact Research, 14437 Potsdam, Germany, and the Department of Physics, Humboldt University of Berlin, Berlin, Germany. }
\thanks{Color versions of one or more figures in this article are available at
 https://doi.org/10.1109/TSMC.2024.3389095.}
\thanks{Digital Object Identifier 10.1109/TSMC.2024.3389095}}

\markboth{IEEE TRANSACTIONS ON SYSTEMS, MAN, AND CYBERNETICS: SYSTEMS. }%
{Feng et al.}

\maketitle

\begin{abstract}
Dynamic processes in complex networks are crucial for better
understanding collective behavior in human societies, biological
systems, and the internet. In this paper, we first focus on the
continuous Markov-based modeling of evolving networks with the
birth-death of individuals. A new individual arrives at the group by
the Poisson process, while new links are established in the network
through either uniform connection or preferential attachment.
Moreover, an existing individual has a limited lifespan before leaving
the network. We determine stationary topological properties of these
networks, including their size and mean degree. To address the effect
of the birth-death evolution, we further study the information
dynamics in the proposed network model from the random drift and
natural selection perspective, based on assumptions of
total-stochastic and fitness-driven evolution, respectively. In
simulations, we analyze the fixation probability of individual
information and find that means of new connections affect the random
drift process but do not affect the natural selection process.

\end{abstract}

%
\IEEEpeerreviewmaketitle

\section{Introduction}
The study of social dynamic information models has been widespread in efforts to understand how interactions among individuals contribute to the evolution of population information \cite{li2013non}. Modeling the information dynamic process is essential for analyzing information spreading patterns, preventing virus transmission, controlling rumors, and more \cite{zhang2016dynamics}. The linear threshold model, originally proposed for binary decision-making in economics and sociology \cite{granovetter1978threshold}, has been applied in various directions of complex systems. Based on this, Dodds et al. generalized the threshold model of contagion by explicitly incorporating the memory of past exposures \cite{dodds2004universal}. Inspired by the theory of interacting particle systems \cite{liggett1985interacting}, Goldenberg et al. proposed a cascading model to investigate the marketing issues \cite{goldenberg2001using}, and then motivated numerous works on generalized cascading information propagation models \cite{gruhl2004information}, \cite{saito2009learning}. Another type of widely used model is established on the epidemic propagation models \cite{kermack1927contribution}, \cite{kermack1932contributions}, where each individual can be in several different states. In some studies, these states includes the susceptible state (the individual is not aware of certain information), the infected state (the individual is aware of this information), and the recovery state (the individual is aware of but does not care about it). Different combinations of these states present various information dynamic models that can reveal social features \cite{de2018fundamentals}.

Information dynamic models assume that individuals in the population interact with one another and obtain information from each other. While individuals can be uniformly mixed, a more common and realistic approach is to describe individuals and their interaction relationships as vertices and edges in graph theory \cite{li2021protection}, \cite{de2016spectral}. The end of the last century saw the rise of complex network theory \cite{albert2002statistical}, which provided an effective and inspiring framework for the study of information dynamics \cite{kirst2016dynamic}. Small-world networks were among the pioneering works that provided highly clustered and homogeneous topology properties by rewiring regular networks to introduce increasing amounts of disorder \cite{watts1998collective, barrat2000properties}. Later, Barab\'{a}si and Albert identified common features of growth and preferential attachment in many large networks \cite{barabasi1999emergence}, \cite{barabasi1999mean}, which reproduced the scale-free power-law distribution and indicated that the development of large networks is governed by robust self-organizing phenomena. These works have been inspiring abundant complex network models to explain various sociological and economic phenomena \cite{boccaletti2006complex}, thus promoting the research of information dynamics in complex networks \cite{wang2019coevolution}. Recently, many special network types, including evolving networks \cite{fogelman2008evolving, li2023evolving}, temporal networks \cite{holme2015modern,zeng2023temporal}, and higher-order networks \cite{battiston2020networks}, are getting attention because of their effectiveness and convenience in describing the properties in real complex systems. 

Recently, some studies have noticed that growth is not the unique property of network sizes in some real situations, and there is a phenomenon of node demise in these networks \cite{zhang2016random}, \cite{9712864}. Based on the Markov chain theory, network models with node demise have been theoretically established and confirmed to be effective in fitting real social population data sets \cite{feng2018evolving}. Besides, inspired by the above network models, Zeng et al. \cite{zeng2022spatial} studied the cooperation dynamics with the birth-death of vertices, finding that the replacement of individuals is important for the emergence of cooperative behaviors in some cases. 

To the best of our knowledge, the degree property of the evolving birth-death network is inadequate, which is crucial to understanding the evolving population's stationary structure. Additionally, the dynamic processes of the mentioned birth-death network are not fully explored, especially the fixation of collective behavior. By studying the information dynamics in populations with both increasing and decreasing individuals, we can gain a better understanding of the evolution of information and culture over time \cite{derex2020cumulative}. In this paper, we aim to investigate the effects of individuals' birth-death on information dissemination. Similar to using disease transmission theory to study information dynamics, we propose a random drift model to analyze the spread of particular information. In this model, the information spread process is completely stochastic, and an aware individual spreads the information to all of their neighbors with the same probability. Inspired by previous research on information dynamics using evolutionary game theory \cite{wells2020prosocial}, we further propose a natural selection model to describe the fitness competition between conflicting information. These conflicting information types are considered to be different strategies in the classical evolutionary game theory \cite{nowak1992evolutionary}, \cite{su2019evolutionary}. In this scenario, individuals are more likely to accept information that benefits their fitness in the population. 

It is worth noting that the idea of introducing propagation theory and evolutionary games to information dynamics is not new. For example, Ye et al. \cite{ye2021game} introduced the evolutionary game method to the collective decision dynamics in the epidemic spreading, offering a unified framework to model and predict complex emergent phenomena, including successful collective responses, periodic oscillations, and resurgent epidemic outbreaks. To demonstrate the criticality and consensus formation, Turalska et al. \cite{turalska2014imitation} introduced the spatial evolutionary game theory to determine in an objective fashion the cooperative or anticooperative strategy adopted by individuals. Additionally, Etesami et al. \cite{etesami2015game} studied the asynchronous Hegselmann-Krause model for opinion dynamics from a novel game-theoretic approach and show that the evolution of an asynchronous Hegselmann-Krause model is equivalent to a sequence of best response updates in a well-designed potential game. To study the disinformation propagation on opinion dynamics, Guo et al. \cite{guo2022effect} presents an opinion framework formulated by repeated, incomplete information games that model online social network users' subjective opinions. However, current works on population information dynamics rarely consider the birth-death of individuals. The introduction of both the birth-death of individuals helps to understand the information dynamics over a long-range time with the population succession, which differs from the mentioned and traditional models. Besides, most studies on dynamic processes stress the importance of network structures and dynamic mechanisms, where static and temporal networks, etc. are mainly considered. Although there are articles focusing on the growing network structure along with the dynamic processes \cite{li2021evolutionary}, \cite{demirel2017dynamics}, few studies pay attention to both the birth-death of individuals in the social dynamics, which is a common phenomenon in all populations and leads to group alternation. Studying this topic helps to understand the effects of evolving birth-death on the population and information dynamics. This manuscript establishes the model of this phenomenon and carried out some fresh results on network structures and dynamic processes, along with some open problems to solve.

The main contributions of this article can be listed as follows:

\begin{enumerate}
    \item We supplement the theory of evolving birth-death networks and explore the information dynamic processes from random drift and natural selection (fitness-driven) perspectives, also interpreted as contagion and evolutionary game processes. 
    \item Theorems about the network size and degree of the evolving birth-death network are proven to be stable via stochastic methods, which are verified by detailed simulations and repeated tests. 
    \item Absorptivity of the population information dynamics is proven in the introduced evolving birth-death network model. Based on this, conclusions about fixation probability are given by simulation in the evolving network. Simulations on real-world network data sets are discussed. 
\end{enumerate}

This paper is organized as follows: We introduce our model and present some theoretical results on the evolving network model and the information dynamical model in Secs. \ref{sec: evolving network model} and \ref{sec: opinion diffusion dynamics in the evolving networks} respectively. Simulation results are presented in Sec. \ref{sec: simulations}. Discussions, conclusions, and outlooks are given in Sec. \ref{sec: conclusions and outlooks}. 
\section{Information Dynamics in Evolving Networks}

As is well known, current works on diffusion dynamics are mostly based on a static and growing population but ignore the decrease of individuals in real society for simplicity. However, population succession is a common phenomenon in the evolution of humans, which influences the information form. Consider the internet, instead of the common assumption of static populations in previous studies, there is both the birth of new individuals and the death of current individuals \cite{feng2018evolving}. The same phenomena appear in other groups as well \cite{ho2018birth}, \cite{crawford2012transition}. This presents an open issue for the network modeling and dynamic process analysis for the evolving birth-death network because the evolution of the population is essential in the long-range time dynamics. Therefore, we here develop a model to study the information dynamics with both the birth-death of individuals. 

In this section, we introduce our proposed model and present some theoretical analysis. First, we describe the evolving network model based on the birth-death process and its general information diffusion dynamics. Next, we establish the random drift information diffusion dynamics in the evolving network model, considering that the propagation process is stochastic and equivalent. Finally, we present the natural selection information diffusion dynamics, where information with high fitness is more likely to spread extensively. 
\subsection{Evolving Network Model}
\label{sec: evolving network model}
 Our employed notations in Sec. \ref{sec: evolving network model} are given in Tab. \ref{tab: notations}. 
\begin{table}[h]
\centering
\caption{Notations}\label{tab: notations}
\begin{tabular}{cc}
   \toprule
   Symbol & Definition\\
   \midrule
   $\lambda$ & The birth rate of the population\\
   $\mu$ & The death rate of an arbitrary current individual\\
   $m$ & The new individual connection number\\
   $N(t)$ & The evolving network size\\
   $p_{i,j}(\Delta t)$ & The network size transition probability from $i$ to $j$ \\
   $q_{i,j}$ & The the network size transition rate from $i$ to $j$\\
   $p_{k(i,j)}(\Delta t)$ & The individual degree transition probability from $i$ to $j$ \\
   $q_{k(i,j)}$ & The individual degree transition rate from $i$ to $j$ \\
   \bottomrule
\end{tabular}
\end{table}
Define an unweighted and undirected graph at time $t$ as $G(t) = (V(t), E(t))$, where $V(t)$ and $E(t)$ denote the vertex set and edge set at time $t$ respectively. The network evolving process starts at $t=0$, where the graph is denoted as $G(0)$ with the vertex set $V(0)$ and the edge set $E(0)$. We define $N(t)=|V(t)|$ as the number of individuals at the time $t$ in the evolving networked population. It is obvious that $\{N(t), t\geq 0\}$ is a stochastic process with the infinite state space $\Omega=\{0,1,2,\cdots\}$. 

The proposed evolving network model is based on the birth-death process (BDEN). We assume that the occurrence of individuals' birth-death follows a Poisson statistics because of both its successful applications in previous studies (e.g. \cite{mcclean1976continuous}, \cite{mei2011early}) and the model simplicity. On the continuous time axis, new individuals arrive at the networked population by the Poisson process with the positive rate $\lambda$ (also interpreted as the birth rate of the system) and connect to $m$ current individuals randomly. That is, during $\Delta t$, there are $k$ new individuals arriving at the system with the probability given by the Poisson distribution denoted as 
\begin{small}
\begin{equation}
    p(k)=\frac{e^{-\lambda \Delta t}(\lambda \Delta t)^k}{k!}.
\end{equation}
\end{small}

Considering that the individual in real systems attaches to current vertices in different means, we assume a new individual connects to existing vertices in two independent ways, including uniform connection and preferential attachment. For the uniform connection (BDEN-uc), the focused individual connects to $m$ current vertices with equal probability. For the preferential attachment (BDEN-pa), it connects to $m$ vertices with the probability proportional to the vertex degree. Besides, each current individual in the population leaves the system with the positive rate $\mu$ (also interpreted as each individual's leave rate of the system), i.e., each existing individual stays in the networked population for a time duration that follows the exponential distribution
\begin{small}
\begin{equation}
    F(t)=1-\exp{(-\mu t)}.
\end{equation}
\end{small}

\begin{figure}                                              
\centering
\includegraphics[scale=0.07]{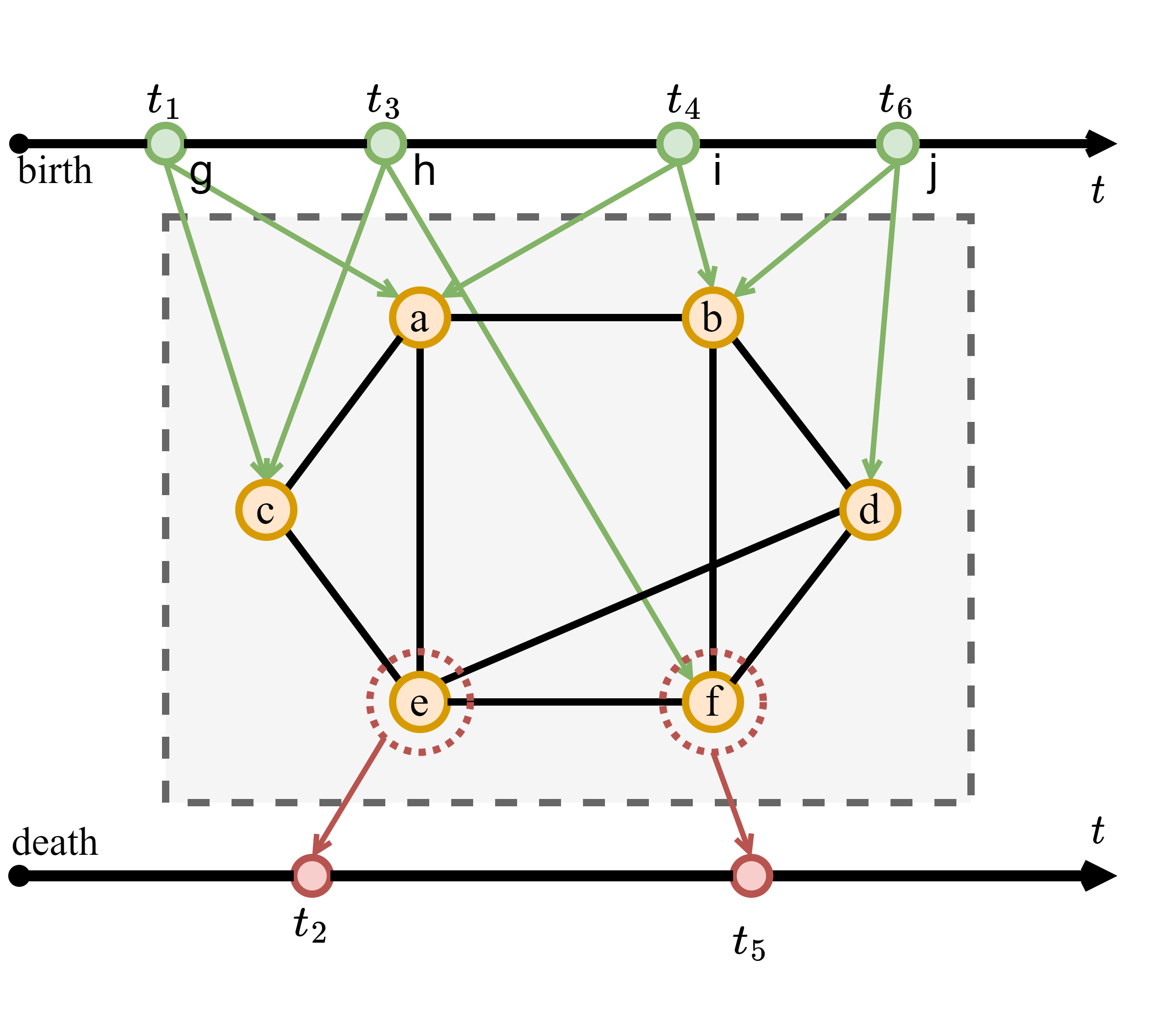}
\vspace{-8mm}
\caption{\textbf{An example of the evolving network model. }This figure shows an example of the evolving networked population based on the birth-death process. The vertices in the initial network are in orange, where the corresponding vertex set is $V(0)=\{a, b, c, d, e, f\}$. New individuals, including $g, h, i, j$, arrive at the network and connect to the existing vertices by Poisson process at $t_1, t_3, t_4, t_6$ respectively. At $t_2$ and $t_5$, the vertices $e$ and $f$ leave the system and disconnect all their neighbors. For instance, the vertex set and edge set at $t_2$ are $V(t_2)=\{a, b, c, d, f, g\}$ and $E(t_2)=\{ab, ac, ag, bf, bd, cg, df\}$. (Color online)}\label{fig: 1}
\end{figure}

Once each individual's lifespan ends, the vertex and its connection are removed from the evolving network. Accordingly, the number of vertices in the evolving network can be regard as an $M/M/\infty$ queuing system. The first letter $M$ represents the characteristics of the arrival process, suggesting that the stochastic process is a memoryless Poisson process. The second letter $M$ indicates the exponential probability distribution of the waiting time. The last symbol $\infty$ means that the capacity of the networked population is infinite. The evolving network model is shown schematically in Fig. \ref{fig: 1}. Before we perform further theoretical analysis, we give some definitions for a better presentation. 
\begin{Definition}\label{def: 1}
Let $p_{i,j}(\Delta t)$ be the network size ($N(t)$) transition probability from $i$ to $j$ during $\Delta t$, denoted as 
\begin{small}
\begin{equation}
    p_{i,j}(\Delta t)=p(N(t+\Delta t)=j\vert N(t)=i).
\end{equation}
\end{small}
\end{Definition}

\begin{Definition}\label{def: 2}
Let $q_{i,j}$ be the size transition speed from $i$ to $j$ during $\Delta t$, expressed as 
\begin{small}
\begin{equation}
    q_{i,j}=
    \left\{  
    \begin{array}{rcl}
    \lim_{t\rightarrow 0^+}\frac{p_{i,j}(t)}{t} & i\neq j\\
    \lim_{t\rightarrow 0^+}\frac{1-p_{i,j}(t)}{t}& i=j\\
    \end{array}
\right. 
\end{equation}
\end{small}
\end{Definition}
Based on the definitions above, we next study the properties of the evolving network based on the birth-death process, including the network sizes and the mean degrees. In fact, we have the following theorems. 
\begin{Theorem}\label{thm: 1}
If the birth rate and each individual's leave rate of an evolving networked population are $\lambda$ and $\mu$ separately, for both BDEN-uc and BDEN-pa, the expected size of the evolving networked population is 
\begin{small}
\begin{equation}\label{eq: thm1}
E[N]=\frac{\lambda}{\mu}. 
\end{equation}
\end{small}
\end{Theorem}
\begin{Proof}
Obviously, the new connection mechanism does not affect the property of the evolving network size. Therefore, no distinction is made in the following analysis. 

Consider there are $i$ individuals now in the networked population and a short time $\Delta t$ that only allows more than one individual to join in or leave the network with the probability $o(\Delta t)$ that tends to zero. 

According to the definition of the Poisson process, the probability that an individual joins in the population is $\lambda\Delta te^{-\lambda\Delta t}$. The probability that it stays in the network is $e^{-\mu\Delta t}$. Therefore, the probability that the network size is increased by 1 is 
\begin{small}
\begin{equation}
\begin{aligned}
    p_{i,i+1}(\Delta t)&=\lambda \Delta te^{-\lambda \Delta t}\times e^{-\mu\Delta t}
    =\lambda \Delta t+o(\Delta t).\\
\end{aligned}
\end{equation}
\end{small}

The probability that no individual arrives at the system is $e^{-\lambda\Delta t}$. An existing individual leaves the system with a probability given by the binomial distribution denoted as $C_i^1 (1-e^{-\mu\Delta t})(e^{-\mu\Delta t})^{i-1}$. Therefore, the probability that one individual leaves the system during $\Delta t$ is 
\begin{small}
\begin{equation}
\begin{aligned}
    p_{i,i-1}(\Delta t)&=e^{-\lambda\Delta t}\times C_i^1 (1-e^{-\mu\Delta t})(e^{-\mu\Delta t})^{i-1}\\
    &=i\mu\Delta t+o(\Delta t).\\
\end{aligned}
\end{equation}
\end{small}

Accordingly, the probability that the network size remains unchanged is indicated as 
\begin{small}
\begin{equation}
\begin{aligned}
    p_{i,i}(\Delta t)&=1-p_{i,i+1}(\Delta t)-p_{i,i-1}(\Delta t)\\
    &=1-\lambda\Delta t-i\mu\Delta t+o(\Delta t).\\
\end{aligned}
\end{equation}
\end{small}

Additionally, the change of the network size is greater than or equal to 2 during $\Delta t$ with the probability
\begin{small}
\begin{equation}
    p_{i,j}(\Delta t)=o(\Delta t), \vert i-j\vert\geq 2. 
\end{equation}
\end{small}

It is worth noting that the transition probability is only related to the time interval $\Delta t$ but not to the initial time $t$. Therefore, the stochastic process $N(t)$ is a homogeneous Markov process. Besides, it is obvious that each state of the process $N(t)$ is accessible, leading to an irreducible Markov chain. Consider an  extreme short time $\Delta t\rightarrow 0^+$, we get the limit 
\begin{small}
\begin{equation}
    \lim_{\Delta t\rightarrow 0^+}p_{i,j}(\Delta t)=0. 
\end{equation}
\end{small}
Consequently, the stochastic process $N(t)$ is continuous. In conclusion, the homogeneity, irreducibility, and continuity of the stochastic process $N(t)$ hold. Then, we can further study the stability of $N(t)$. 

Based on Def. \ref{def: 2} and the Kolmogorov's differential equations, we have
\begin{small}
\begin{equation}\label{eq: kolmogrov}
    \frac{d p_{i,j}(\Delta t)}{d \Delta t}=-p_{i,j}(\Delta t)q_{j,j}+\sum_{k \in \Omega,k \neq j}p_{i,k}(\Delta t)q_{k,j}.
\end{equation}
\end{small}
Note that $p_{i,j}'(\Delta t)$ converges to 0 because $p_{i,j}(\Delta t)$ is a bounded function with values between 0 and 1. Assume $p_j=\lim_{t\rightarrow\infty}p_{i,j}(t)$ is the limit probability that $N(t)=j$, we have the following result by letting Eq. \ref{eq: kolmogrov} be $0$
\begin{small}
\begin{equation}\label{eq: pi}
    p_i=\frac{\lambda^i}{\mu^i i!}\exp{(-\lambda/\mu)}.
\end{equation}
\end{small}
Then, the expected network size is 
\begin{small}
\begin{equation}
    E[N]=\sum_{i=0}^{\infty}ip_i=\frac{\lambda}{\mu}.
\end{equation}
\end{small}

This ends the proof.
$\hfill\blacksquare$
\end{Proof}

According to Thm. \ref{thm: 1}, we obtain the expected network size in the evolving process of the population if the system's birth rate $\lambda$ and each individual's leave rate $\mu$ are given. The expectation of the network size is a positive number. However, the network size that we can observe in the population is a positive integer. We usually hope that the size of a networked population is greater than zero and valuable for the study of dynamic issues. Therefore, we present the following remark. 
\begin{Remark}
We can set $\lambda>\mu$ to ensure the population size is non-zero for most of the evolving process, and $\lambda\gg\mu$ to ensure the population is large enough for the study of the dynamical process. 
\end{Remark}

To pave the way for the theoretical analysis of the information dynamics in the evolving network model, we now present the following theorem on the expected mean degree of the evolving networked population by using one vertex $v\in V(t)$ to approximate the whole network. Note that $k(t)$ is the stochastic process of the vertex $v$'s degree with the state space $\Omega_k=\{0,1,2,\cdots\}$. 
\begin{Definition}\label{def: 3}
Let $p_{k(i,j)}(\Delta t)$ be the the vertex $v$'s degree ($k(t)$) transition probability from $i$ to $j$ during $\Delta t$, i.e.,
\begin{small}
\begin{equation}
    p_{k(i,j)}(\Delta t)=p(k(t+\Delta t)=j\vert k(t)=i). 
\end{equation}
\end{small}
\end{Definition}

\begin{Definition}\label{def: 4}
Let $q_{k(i,j)}$ be the vertex $v$'s degree ($k(t)$) transition speed that from $i$ to $j$ during $\Delta t$, that is
\begin{small}
\begin{equation}
    q_{k(i,j)}=
    \left\{  
    \begin{array}{rcl}
    \lim_{t\rightarrow 0^+}\frac{p_{k(i,j)}(t)}{t} & i\neq j\\
    \lim_{t\rightarrow 0^+}\frac{1-p_{k(i,j)}(t)}{t}& i=j\\
    \end{array}
\right. 
\end{equation}
\end{small}
\end{Definition}

\begin{Theorem}\label{thm: 2}
If the birth rate and each individual's leave rate of an evolving networked population are $\lambda$ and $\mu$ separately, and each new individual connects to $m$ existing individuals when it joins in the population. Then, the expected degree of each individual in BDEN-uc is independent of both $\lambda$ and $\mu$ and denoted as
\begin{small}
\begin{equation}
    E[k]=m.
\end{equation}
\end{small}
\end{Theorem}
\begin{Proof}
Consider an existing vertex $v\in V(t)$ in the population, during $\Delta t$, its degree increases or decreases by 1 if and only if a new vertex connects to the node $v$ or one of $v$'s neighbors disconnects to the network. Employing $E[N]=\lambda/\mu$ to approximate the network size $N(t)$ at time $t$ according to Thm. \ref{thm: 1}, the probability that the node $v$'s degree increases by 1 is indicated as 
\begin{small}
\begin{equation}
    p_{k(i,i+1)}(\Delta t)=\lambda \Delta t\times \frac{m}{E[N]}+o(\Delta t). 
\end{equation}
\end{small}

The probability that a new individual arrives at the system but does not connect to the individual $v$ is $\lambda\Delta te^{-\lambda\Delta t}(1-\frac{m}{E[N]})$. No individual arrives at the population during $\Delta t$ with the probability $e^{-\lambda\Delta t}$. Besides, one of the vertex $v$'s neighbors leaves the system with the probability given by the binomial distribution $C_i^1 (1-e^{-\mu\Delta t})(e^{-\mu\Delta t})^{i-1}$ if there are $i$ individuals in the population. Accordingly, the probability that the node $v$'s degree decreases by 1 is 
\begin{small}
\begin{equation}
\begin{aligned}
    p_{k(i,i-1)}(\Delta t)&=(\lambda\Delta te^{-\lambda\Delta t}(1-\frac{m}{E[N]})+e^{-\lambda\Delta t})\\
    &\times C_i^1 (1-e^{-\mu\Delta t})(e^{-\mu\Delta t})^{i-1}\\
    &=i\mu\Delta t+o(\Delta t)\\
\end{aligned}
\end{equation}
\end{small}
Therefore, the vertex $v$'s degree stays unchanged during $\Delta t$ with the probability
\begin{small}
\begin{equation}
\begin{aligned}
    p_{k(i,i)}(\Delta t)&=1-p_{k(i,i+1)}(\Delta t)-p_{k(i,i-1)}(\Delta t)\\
    &=1-\lambda \Delta t\times \frac{m}{E[N]}-i\mu\Delta t+o(i\mu\Delta t)\\
\end{aligned}
\end{equation}
\end{small}
Next, the probability that the vertex $v$'s degree changes greater than 2 during $\Delta t$ is 
\begin{small}
\begin{equation}
    p_{k(i,j)}=o(\Delta t), \vert i - j \vert \geq 2.
\end{equation}
\end{small}
Denote $p_{k(i)}$ as the limit probability that the vertex $v$'s degree is $i$ provided $t\rightarrow\infty$. When it comes to the stable state of the stochastic process $\{k(t),t\geq0\}$, the speed at which the process enters or exits a state is the same. Therefore, we have

\begin{small}
\begin{equation}
\frac{m}{E[N]}\lambda p_0 =p_1 \mu 
\end{equation}
\end{small}
and

\begin{small}
\begin{equation}
p_{i-1}\frac{m}{E[N]}\lambda+p_{i+1}(i+1)\mu=p_i(\frac{m}{E[N]}\lambda+i\mu), i=1,2,\cdots 
\end{equation}
\end{small}
Solving the system of equations above, we obtain the limit probabilities
\begin{small}
\begin{equation}
    p_0=(1+\sum_{j=1}^{\infty}\frac{1}{j!}(\frac{m\lambda}{E[N]\mu})^j)^{-1}=\exp{(-\frac{m\lambda}{E[N]\mu})},
\end{equation}
\end{small}
and
\begin{small}
\begin{equation}
    p_i=\frac{1}{i!}(\frac{m\lambda}{E[N]\mu})^ip_0, i=1,2,\cdots. 
\end{equation}
\end{small}
Therefore, the expected degree of the vertex $v$ is 
\begin{small}
\begin{equation}
\begin{aligned}
    E[k]&=\sum_{i=0}^\infty ip_i\\
    &=\frac{m\lambda}{E[N]\mu}\exp{(-\frac{m\lambda}{E[N]\mu})}(\sum_{i=0}^\infty \frac{1}{i!}(\frac{m\lambda}{E[N]\mu})^i)\\
    &=m.\\
\end{aligned}
\end{equation}
\end{small}
The last equal sign in the above formula holds because of the Thm. \ref{thm: 1}. 

The result follows. 
$\hfill\blacksquare$
\end{Proof}

Thm. \ref{thm: 2} shows that whatever the positive numbers $\lambda$ and $\mu$ are, the expected degree of the BDEN-uc is $m$. Additionally, for BDEN-pa, this mean-field deduction is not applicable since the heterogeneity of new connections. This ends the modeling part of the evolving networked population based on the birth-death process (Sec. \ref{sec: evolving network model}). 
\subsection{Information Dynamics in the Evolving Networks}
\label{sec: opinion diffusion dynamics in the evolving networks}

\begin{figure}                                              
\centering
\includegraphics[scale=0.05]{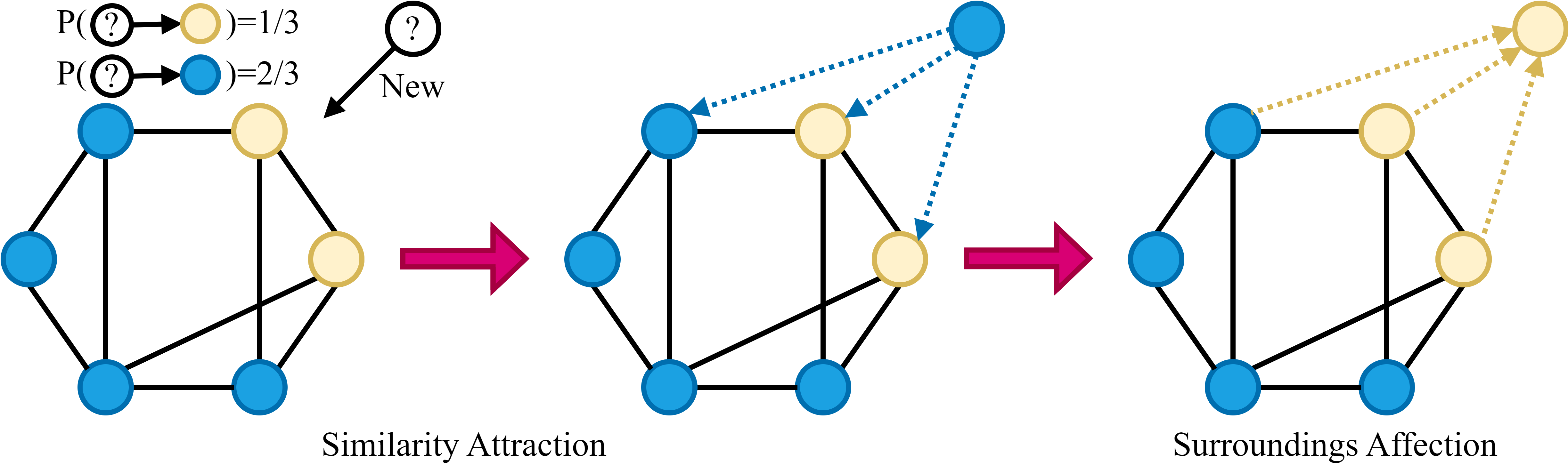}
\vspace{-2mm}
\caption{\textbf{An example of the information dynamic model. }This figure shows an example of the information diffusion process on the mentioned evolving network model. Vertices in blue and yellow denote the two states among the population. Here, when a new individual is arriving, it plays the same strategy as the blue nodes with the probability $2/3$, meaning that newcomers are more likely to be attracted to the population with the mainstream information in our model. After connecting to existing vertices, this new individual is affected by surroundings, which helps to adapt to the environment. (Color online)}\label{fig: 2}
\end{figure}

In this subsection, inspired by both the propagation theory and the evolutionary game theory, we propose our information diffusion dynamical mechanism in the evolving networked population above. Our mainly employed notations in Sec. \ref{sec: opinion diffusion dynamics in the evolving networks} are given in Tab. \ref{tab: notations2}. 
\begin{table}[h]
\centering
\caption{Notations}\label{tab: notations2}
\begin{tabular}{cc}
   \toprule
   Symbol & Definition\\
   \midrule
   $A(t)$ & The number of aware individuals in random drift at $t$\\
   $U(t)$ & The number of unaware individuals in random drift at $t$\\
   $C(t)$ & The number of $C$ individuals in natural selection at $t$\\
   $D(t)$ & The number of $D$ individuals in natural selection at $t$\\
   $\alpha$ & The transmission rate of an aware individual in random drift\\
   $M$ & The payoff matrix of the natural selection\\
   $\delta$ & The intensity of natural selection\\
   $f_x$ & The fitness of an $x$ individual\\
   \bottomrule
\end{tabular}
\end{table}

The dynamic process of network population has been widely studied during the past two decades. Most dynamic studies in evolving or temporal networks focus on the proportion of a specific behavior \cite{zeng2022spatial2,li2021protection, feng2023evolutionary}, which neglects the fixation probability of the behavior in a finite population. Here, we analyze the collective behavior by fixation probability of the evolving network from two perspectives, including random drift and natural selection. 

\textit{The random drift process} assumes that the information is spreading randomly from the individual that is aware of the information interpreted as the state $A$ to the unaware individual interpreted as the state $U$ in the whole population. Once an individual is aware of this information, it will not become unaware. Additionally, an unaware individual can be infected by aware individuals among its neighbors. 

\textit{The natural selection process} assumes that two kinds of incompatible information are conflicting, the individuals can be aware of both the information $C$ and a different information $D$. It is worth noting that the states $C$ and $D$ are two kinds of conflicting information in the whole population. In this case, the individuals are assumed to be more likely to hold the information with a high fitness among their surroundings. The natural selection process is game-theoretic and fitness-driven. 

Before examining the specific dynamical mechanisms, we introduce a general information diffusion model. In real society, individuals who are similar to the population are more likely to be accepted, which is interpreted as similarity attraction in our model. Therefore, we assume that a new individual holds certain information with a probability equal to the proportion of that information in the network before their arrival. Then, the new individual connects to $m$ existing individuals and is influenced by their surroundings via the random drift process or the natural selection process. In conclusion, we put forward that when a new individual arrives at the network, its general diffusion dynamics can be described as follows, and an example is provided in Fig. \ref{fig: 2}. 

\textbf{Similarity Attraction. }Before the new individual constructs a relationship with other individuals in the population, it becomes aware of information with a probability equal to the frequency of this information, which is considered as its initial information; 

\textbf{Surroundings Affection. }After the initial information is determined, the individual connects randomly to $m$ existing vertices and changes its information according to its surrounding neighbors by the random drift process or the natural selection process. 


Then, we introduce the specific concept of the random drift process and the natural selection process. 
\subsubsection{Random Drift}
\label{sec: random drift}
Consider a specific information is spreading in the networked population. There are two states for these individuals on the information, including being aware $A$ and unaware $U$ of this information. An aware individual can spread its information to an unaware individual with the probability $\alpha$. Additionally, once an individual becomes aware of the information, it will not turn unaware again. Let $\{A(t),t\geq 0\}$ and $\{U(t),t\geq 0\}$ be the stochastic processes of the number of aware and unaware individuals separately with the state space $\Omega_A=\Omega_U=\{0,1,2,\cdots\}$, where $N(t)=A(t)+U(t)$ for random drift. The random drift information diffusion process of a new individual is as follows:

1. Before an individual arrives at the network and constructs relationships with others, it is aware of the information ($A$) with the probability equal to the ratio of $A$ individuals in the population ($A(t)/N(t)$) or is unaware of the information with the probability $U(t)/N(t)$. 

2. After the individual already connects to $m$ neighbors, each of its aware neighbors spreads the information to the focal individual with the probability $\alpha$. 

\subsubsection{Natural Selection}
\label{sec: nature selection}
We next consider that two kinds of information ($C$ and $D$) are conflicting in the evolving networked population and define the paired vertices with the same information as the consensus state. We establish our model employing the evolutionary game theory. Specifically, suppose an interaction between two individuals, a $C$ individual receives the payoff $R$ if it reaches a consensus with another individual with the same information $C$, and a $D$ player receives the payoff $P$ while interacting with another $D$ holder. When they reach no consensus, the individual with the information $C$ obtains the payoff $S$, and $D$ earns the payoff $T$. Therefore, we can define the payoff matrix for the natural selection information diffusion process as
\begin{small}
\begin{equation}
M=\left[
\begin{array}{cc}
 R & S  \\
 T & P \\
\end{array}
\right].
\end{equation}
\end{small}
Then, if an individual $x$ has the neighbor set $N_x$, the total payoff of the individual $x$ is the sum of payoff from all neighbors and indicated as
\begin{small}
\begin{equation}
\Pi_x=\sum_{y\in N_x}s_x^T M s_y,
\end{equation}
\end{small}
where $s_x=[1,0]$ or $[0,1]$ if $x$ is aware of the information $C$ or $D$ separately. Consequently, we map this payoff function to the fitness function as 
\begin{small}
\begin{equation}
f_x=1-\delta+\delta\Pi_x,
\end{equation}
\end{small}
where $\delta\geq 0$ denotes the intensity of selection. We are
especially concerned with the effects of weak selection,
meaning that $0<\delta\ll 1$. 

Let $\{C(t),t\geq 0\}$ and $\{D(t),t\geq 0\}$ be the stochastic processes with the state spaces $\Omega_C=\Omega_D=\{0,1,2,\cdots\}$ of the number of individuals holding $C$ and $D$ respectively, where $N(t)=C(t)+D(t)$ for natural selection. With the assumptions above, the natural selection information diffusion process can be introduced as follows:

1. Before the individual arrives the network and interacts with others, it holds the information $C$ with the probability $C(t)/N(t)$, or holds the information $D$ with the probability $D(t)/N(t)$. 

2. Once the individual arrives and connects to $m$ neighbors, its neighbors compete to transmit their information to the new individual. The new individual accepts its neighbors' information with the probability proportional to their fitness. 

It is worth noting that this dynamical mechanism is inspired by the death-birth update rule in the spatial evolutionary game theory. 

Additionally, we carry out the following theorem to demonstrate that there are two absorption states in the evolving networked population, including the pure aware and the pure unaware state. Studying the existence of both absorption states helps to explore whether random drift and natural selection can favor certain information in the networked population. 

\begin{Theorem}\label{thm: 3}
If the evolving networked population undergoes the random drift or natural selection information diffusion process, it will converge to one of the pure information absorbing states for both BDEN-uc and BDEN-pa. 
\end{Theorem}
\begin{Proof}
Let us first focus on the stochastic process $\{A(t), t\geq 0\}$ during $\Delta t$. A new individual is aware before it arrives in the system at the time $t$ with the probability $A(t)/N(t)$. The new individual is unaware before the arrival, and it becomes aware when interacting with its neighbors with the probability $[1-(1-\alpha)^{A(t)/N(t)}]\times U(t)/N(t)$. Therefore, for the random drift process, a new individual becomes aware during $\Delta t$, i.e., $A(t)$ increases by 1, with the probability 
\begin{small}
\begin{equation}
\begin{aligned}
    &p(A(t+\Delta t)- A(t)=1)\\
    &=\{\frac{A(t)}{N(t)}+[1-(1-\alpha)^{m\frac{A(t)}{N(t)}}]\frac{U(t)}{N(t)}\}\lambda[\Delta t+o(\Delta t)].\\
\end{aligned}
\end{equation}
\end{small}

The probability that an aware individual leaves the population is
\begin{small}
\begin{equation}
\begin{aligned}
    &p(A(t+\Delta t)-A(t)=-1)\\
    &=C_{A(t)}^1 (1-e^{-\mu\Delta t})(e^{-\mu\Delta t})^{A(t)-1}\\
    &=A(t)[\mu\Delta t+o(\Delta t)].\\
\end{aligned}
\end{equation}
\end{small}

Obviously, at time $t$, for an arbitrary state $A(t)>0$ ($A(t)\in \Omega-\{0\}$), the stochastic process goes to a transient state. After a sufficient time $t$, the process $A(t)$ reaches the absorbing state if $A(t)=0$, which guarantees the number of aware individuals converges to 0. Therefore, $A(t)=0$ is an absorbing state of the networked population if the process evolves for a sufficient time $t$, i.e., 
\begin{small}
\begin{equation}
\lim_{t\rightarrow\infty}A(t)=0.
\end{equation}
\end{small}

Note that if $A(t)=N(t)$, the transition probability of the stochastic process $A(t)$ is the same as $N(t)$. According to our definitions and assumptions, if all individuals are aware, the population does not accommodate the unaware individual. Additionally, we can induce $U(t)=0$ by $A(t)=N(t)$ and $A(t)+U(t)=N(t)$. Therefore, $U(t)=0$ is another absorbing state for a sufficient time $t$, denoted as 
\begin{small}
\begin{equation}
\lim_{t\rightarrow\infty}U(t)=0. 
\end{equation}
\end{small}

Similarly, consider an arbitrary non-pure state, both $C(t)$ and $D(t)$ will evolve into a neighbor state. There are two absorption states, including $C(t)=0$ and $C(t)=N(t)$ for the natural selection process to evolve into, while every other state is transient. 

This ends the proof. 
$\hfill\blacksquare$
\end{Proof}

In our proof for Thm. \ref{thm: 3}, we carry out that the evolving network converges to an absorption state with a sufficient time. Additionally, we have another explanation on Thm. \ref{thm: 3} as follows. 
\begin{Remark}
The stochastic process $\{A(t),t\geq 0\}$ will converge to $A(t)=0$ or to another process $N(t)$, and the stochastic process $\{U(t),t\geq 0\}$ will converge to $U(t)=0$ or to another process $N(t)$. 
\end{Remark}

\begin{Remark}
The stochastic process $\{C(t),t\geq 0\}$ will converge to $C(t)=0$ or to the other process $N(t)$, and the stochastic process $\{D(t),t\geq 0\}$ will converge to $D(t)=0$ or to the other process $N(t)$. 
\end{Remark}

The exact solution of the fixation probability of each information type (the probability that mutant individuals occupy the whole population) is an open issue in this article. We assume if a $C$ individual invades the population, the evolving network favors cooperation if the probability of this focal invader occupying the whole network is higher than $1/E[N]$. Based on the identity-by-descent (IBD) method \cite{allen2014games}, we have the following theorem for the natural selection process in BDEN-uc. 
\begin{Theorem}\label{thm: 4}
For prisoner's dilemma natural selection with $R=b-c$, $S=-c$, $T=b$, and $P=0$, the population favors $C$ if
\begin{equation}
\frac{b}{c}>\frac{\lambda/\mu-2}{\lambda/m\mu-2}. 
\end{equation}
\end{Theorem}
\begin{Proof}
Consider a $C$ individual as our focal vertex with fitness $f_0$, which we say is the vertex 0. Denote $\Pi^{(n)}$ as the payoff from $n$ steps random walk away from the vertex 0, $s^{(n)}$ as the probability that an individual is $C$ from $n$ steps away, $p^{(n)}$ is the probability that an $n$-step random walk ends at where it starts. The IBD method means that the evolving population favors cooperation ($C$) if 
\begin{small}
\begin{equation}\label{eq: IBD}
\left \langle \frac{\partial(b_0-d_0)}{\partial\delta} \right \rangle_{\delta=0, s^{(0)}=1}>0, 
\end{equation}
\end{small}
where $b_0$ and $d_0$ is the probability for the focal $C$ individual to generate a $C$ offspring and be replaced by neighbors respectively. 

We have mentioned that the state transition is proportional to fitness. Therefore, we have 
\begin{small}
\begin{equation}\label{eq: b0}
b_0=\sum_{i\in V}\frac{1}{N}\frac{f_0}{\sum_{l\in V}f_l}, 
\end{equation}
\end{small}
and
\begin{small}
\begin{equation}\label{eq: d0}
d_0=\frac{1}{N}\sum_{j\in V}\frac{f_j}{\sum_{l\in V}f_l}. 
\end{equation}
\end{small}

Substituting Eqs. \ref{eq: b0} and \ref{eq: d0} into Eq. \ref{eq: IBD}, the condition that the system favors $C$ becomes $\Pi^{(0)}-\Pi^{(2)}>0$. The payoff from $n$-step away is
\begin{small}
\begin{equation}
\begin{aligned}
&\Pi^{(n)}=s^{(n)}(s^{(n+1)}(b-c)+(1-s^{(n+1)})(-c))\\
&+(1-s^{(n)})s^{(n+1)}b=-cs^{(n)}+bs^{(n+1)}\\.
\end{aligned}
\end{equation}
\end{small}
Therefore, the condition Eq. \ref{eq: IBD} further becomes
\begin{small}
\begin{equation}\label{eq: IBD2}
-cs^{(0)}+bs^{(1)}+cs^{(2)}-bs^{(3)}>0. 
\end{equation}
\end{small}

Following Refs. \cite{allen2014games} and \cite{wang2023greediness}, $s^{(n)}$ can be transformed by 
\begin{small}
\begin{equation}
s^{(n)}-s^{(n+1)}=\frac{u}{2}(Np^{(n)}-1)+O(u^2), 
\end{equation}
\end{small}
where $u$ denotes the mutation probability, which we do not consider here and can be reduced later, and $N$ can be replaced by our mentioned Thm. \ref{thm: 1} with $E[N]=\lambda/\mu$. 

Accordingly, Eq. \ref{eq: IBD2} becomes 
\begin{small}
\begin{equation}
(2-p^{(0)}\lambda/\mu-p^{(1)}\lambda/\mu)c+(p^{(1)}\lambda/\mu+p^{(2)}\lambda/\mu-2)b>0. 
\end{equation}
\end{small}
Herein, $p^{(0)}=1$ since a random walk must be where it starts without moving. A random walk is impossible to return to the initial place with only one step out, which leads to $p^{(1)}=0$. Based on Thm. \ref{thm: 2}, a two-step random walk returns to the starting vertex with probability $p^{(2)}=m\times \frac{1}{m}\times \frac{1}{m}$. Therefore, the BDEN-uc favors $C$ behavior if 
\begin{small}
\begin{equation}
\frac{b}{c}>\frac{p^{(0)}\lambda/\mu+p^{(1)}\lambda/\mu-2}{p^{(1)}\lambda/\mu+p^{(2)}\lambda/\mu-2}=\frac{\lambda/\mu-2}{\lambda/m\mu-2}. 
\end{equation}
\end{small}
Results follow.
$\hfill\blacksquare$
\end{Proof}
This ends the modeling part of the evolving network based on the birth-death process and the information diffusion dynamics. 

\section{Simulations}
\label{sec: simulations}
\subsection{Methods}
In simulations of the evolving networks, we mainly focus on the network sizes and mean degrees. As stated previously, we need to set the birth rate of the system much larger than each individual's leave rate to ensure that the expected sizes of the evolving networks are large enough for us to analyze the information diffusion dynamics. $\lambda$, $\mu$, and $m$ present the birth rate, the death rate, and the new connection number respectively. $\lambda$ stands for the Poisson rate of the new individuals' arrival, which leads to a linear expectation $\lambda\Delta t$ of newcomers during a time $\Delta t$. Besides, $\mu$ is supposed to be far less than $\lambda$ to ensure a sufficient evolving population size for the study of network topology and information dynamics. Therefore, the main parameters we use are $\lambda$s$=[2,3,4,5]$, $\mu=0.01$, $m$s$=[4,5,6,8,10]$. Accordingly, the evolving network sizes are from $200$ to $500$, which are enough for the study of fixation probability \cite{ohtsuki2006simple}. Additionally, the initial network $G(0)$ is the complete graph with $N(0)=30$ vertices. 

Regarding the generation of the evolving birth-death network, we employ the fact that the inter-event time of a Poisson with the parameter $\lambda$ flow follows the exponential distribution with the parameter $\lambda$ as well. Accordingly, once a vertex is generated, the next vertex is arriving in the following exponential random time. Each generated individual has an exponentially distributed lifespan. Once the lifespan of an individual ends, it is removed from the evolving network. The evolving process of the birth-death network ends until a sufficient time, which is defined differently in the simulation about the network topology and the information dynamics. For the study of network topology, the evolving process reaches sufficient time if the network size and degree are relatively stable around some values. For the study of information dynamics, sufficient time is reached if the networked population evolves into one of the absorption states. 

\begin{figure*}                                              
\centering
\subfigure[Evolution of Network Sizes]{
\includegraphics[scale=0.25]{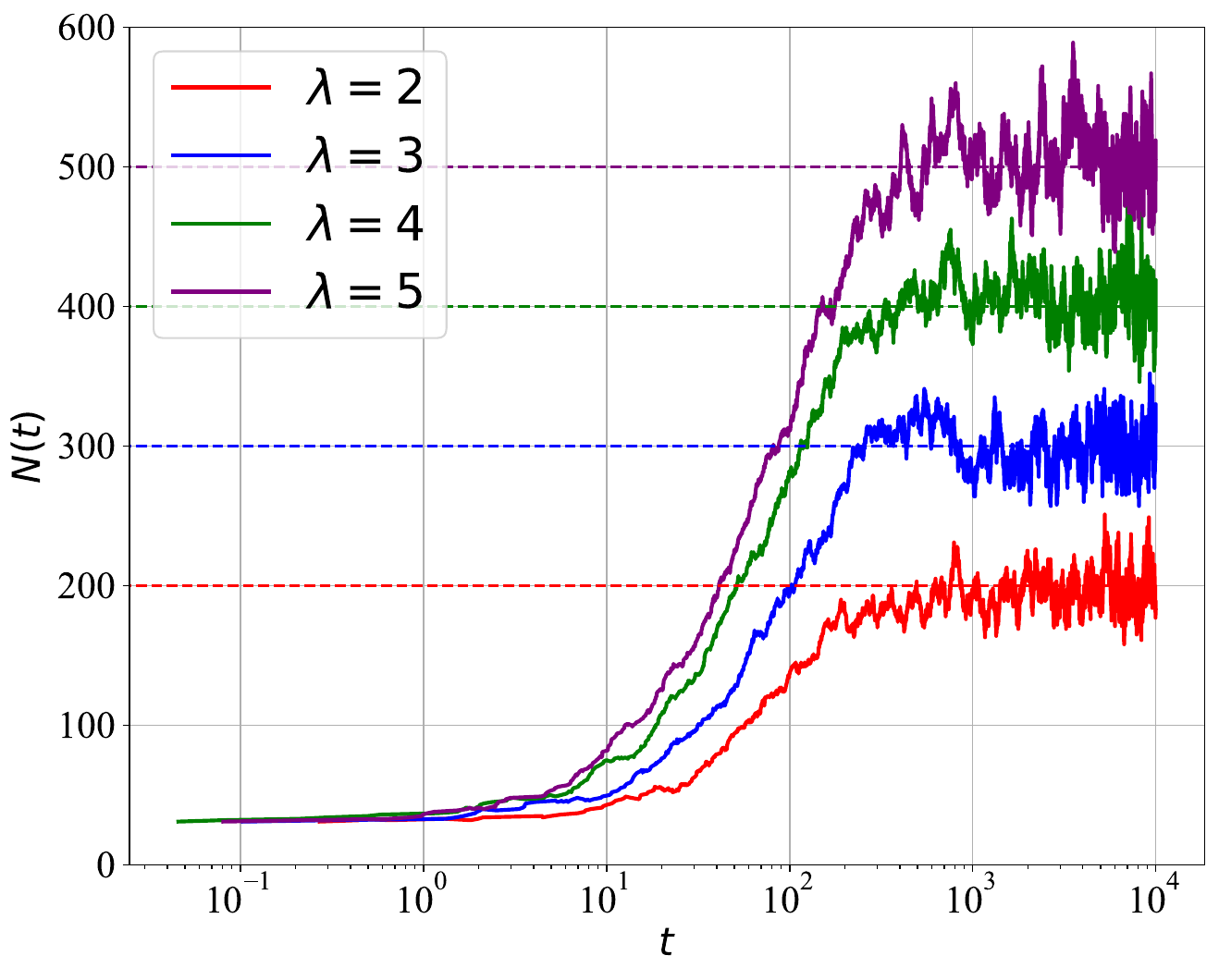}
\label{fig: 4(a)}
}
\subfigure[Expected Network Sizes]{
\includegraphics[scale=0.25]{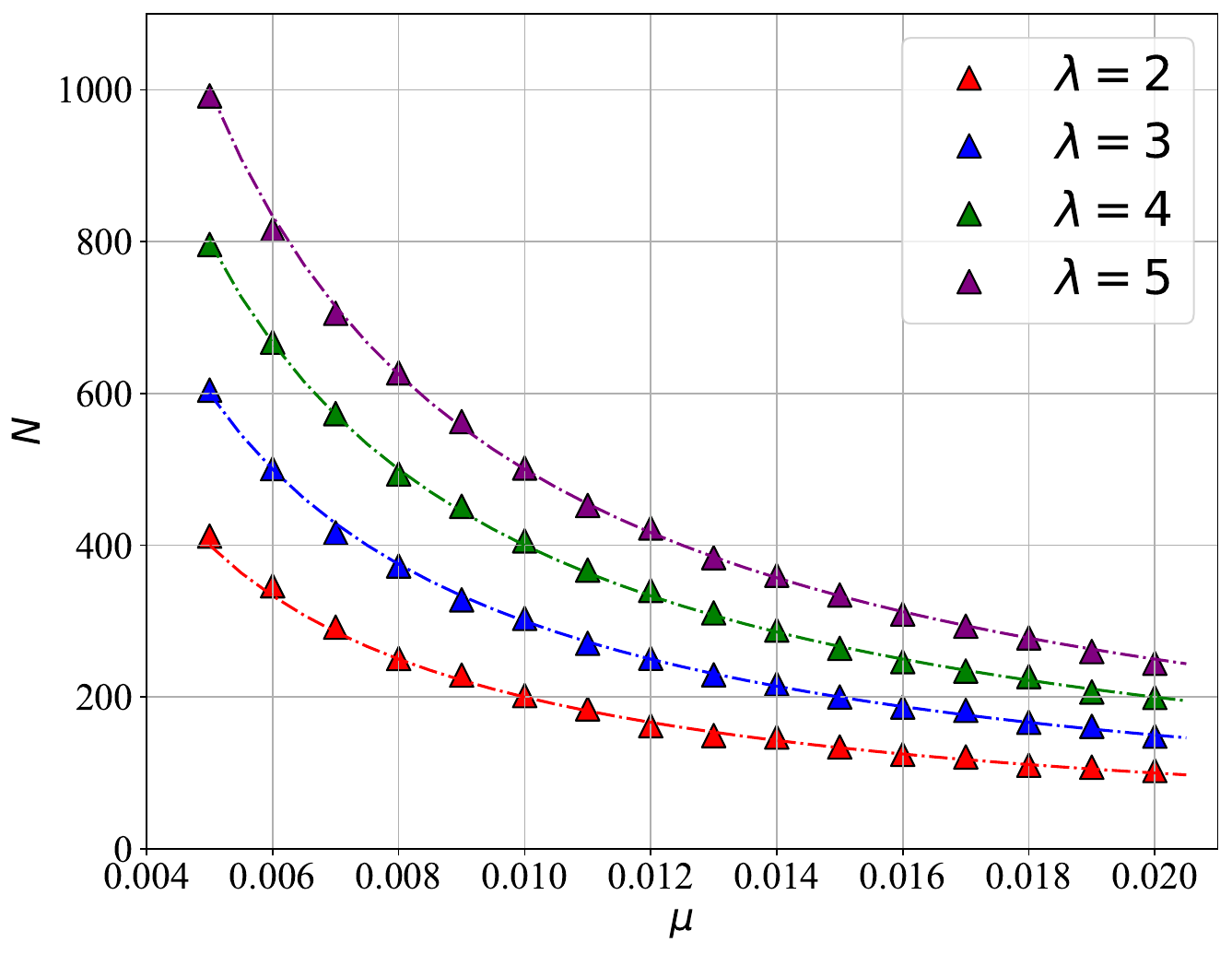}
\label{fig: 4(b)}
}
\subfigure[Frequency of Sizes]{
\includegraphics[scale=0.25]{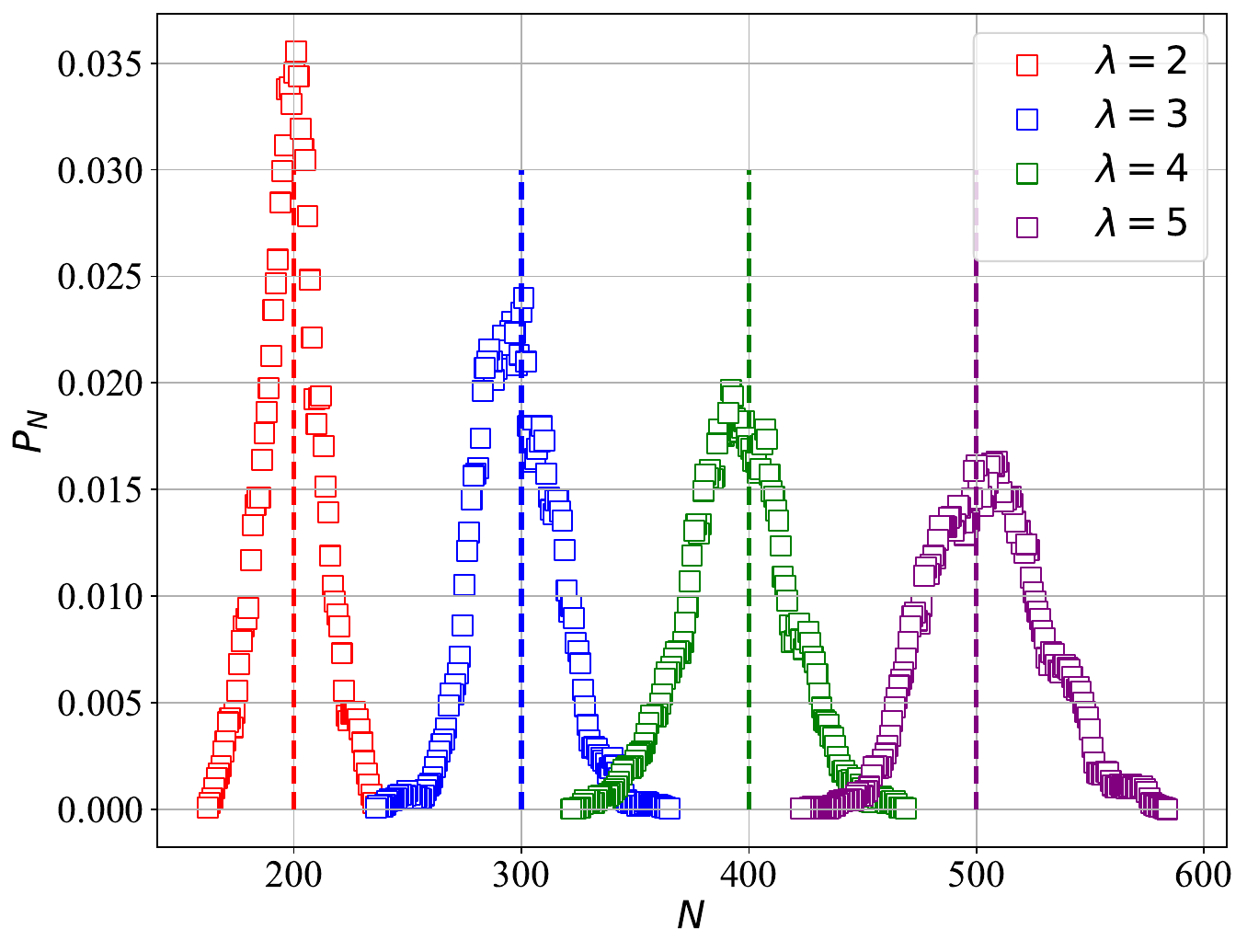}
\label{fig: 4(c)}
}
\caption{\textbf{Results on network sizes. }(a) The evolution of network sizes from $t=0$ to $t=10^4$. (b) The comparison of the theoretical results to the simulations on the expected network sizes. (c) The frequency of network sizes in the evolving process. These results hold for both BDEN-uc and BDEN-pa. Here, we set $\lambda$s$=[2,3,4,5]$, $m=5$, and the initial network size as $N(0)=30$ for all results. In (a), we set $\mu=0.01$. The $x$-axis and $y$-axis are set as the time $t$ and the network size $N(t)$. Dashed lines present the expectation $E[N]=\lambda/\mu$. In (b), the $x$-axis is set as the leave rate $\mu\in[0.005,0.02]$, and the simulations are carried out with the interval $0.001$. The $y$-axis is set as the average network size. Each data point in triangle is obtained by averaging the network sizes from $t=7\times10^3$ to $t=10^4$. Dashed lines are the theoretical solutions of expected network sizes. In (c), we set the $x$-axis as the network size $N$, and the $y$-axis as the corresponding probability in the evolving process. The samples are the network sizes from $t=10^3$ to $t=10^4$. (Color online)}
\label{fig:4}
\end{figure*}

We present our simulations on both the evolving networked population and the information diffusion dynamics. All following results are carried out on \textit{Python 3.8}. We construct the evolving network via the package \textit{networkx} and generate the variables via the package \textit{numpy.random}. Primarily, we present our simulation results on the network sizes and the mean degrees. Then, we introduce and study the fixation probability of both the random drift and natural selection processes. 

\subsection{Network Model Based on the birth-death Process}
Basically, we present the results on network sizes (the stochastic process $\{N(t),t\geq 0\}$) for $\lambda$s$=[2,3,4,5]$ and $\mu=0.01$. Note that the result of network sizes is irrelevant to the connection mechanisms. Therefore, our statement for the network sizes holds for both BDEN-uc and BDEN-pa. As shown in Fig. \ref{fig: 4(a)}, $N(t)$s increase with time $t$ from $N(0)=30$ and then fluctuate around the expectations defined in Eq. \ref{eq: thm1}. With a fixed individual's leave rate $\mu$, a high birth rate of the population ensures that the network size increases rapidly and becomes stable around a high expectation. Besides, the fluctuation is sharp given a high birth rate $\lambda$. Dashed lines in Fig. \ref{fig: 4(b)} present the corresponding theoretical expectations for the four groups of parameters. We find that the theoretical expectations fit well with the evolution stable state. Additionally, the network sizes are already stable around the expectation if $t>10^3$, which provides evidence for us to perform the following analysis. 

By averaging the network sizes from $t=7\times10^3$ to $10^4$, we obtain the average network sizes in Fig. \ref{fig: 4(b)} for $\lambda$s$=[2,3,4,5]$ and $\mu\in[0.005,0.02]$. It is worth noting that the parameter $\lambda$s and $\mu$s can be any positive numbers. Nevertheless, to ensure the results are meaningful, we set each individual's leave rate $\mu$ in this range and perform the simulations with a sampling of $0.001$. Comparing the simulation results in triangles and the theoretical results in dashed lines, we find the Thm. \ref{thm: 1} confirmed. Additionally, the mean network sizes decrease as the increase of each individual's leave rate $\mu$ with a certain fixed birth rate $\lambda$.  However, with a fixed leave rate $\mu$, the expected network size grows with the increase of the birth rate $\lambda$. 

As stated previously, the stochastic process of the network size $N(t)$ fluctuates around the expectation. Then, we pay attention to the deviation degree of $N(t)$ from the expectation. The simulations are performed with $\lambda$s$=[2,3,4,5]$ and $\mu=0.01$. By recording the network sizes of each group of the network parameters from $t=10^3$ to $10^4$, we present the frequency of the network sizes at the evolution stable state. Obviously, the frequency of network sizes is symmetrical and follows the Poisson distribution with the parameter $\lambda/\mu$ as Eq. \ref{eq: pi} suggests. Therefore, a network size is less or more than the expectation with an equal probability. Additionally, for a low birth rate $\lambda$, there is a high probability to observe a network size that is close to the expectation in the evolution process. However, when it comes to a higher $\lambda$, the fluctuation is enhanced, leading to a higher probability for the population to possess a number of individuals that are far from the expectation. 

\begin{figure*}                                              
\centering
\subfigure[Evolution with Fixed $m$ (BDEN-uc)]{
\includegraphics[scale=0.25]{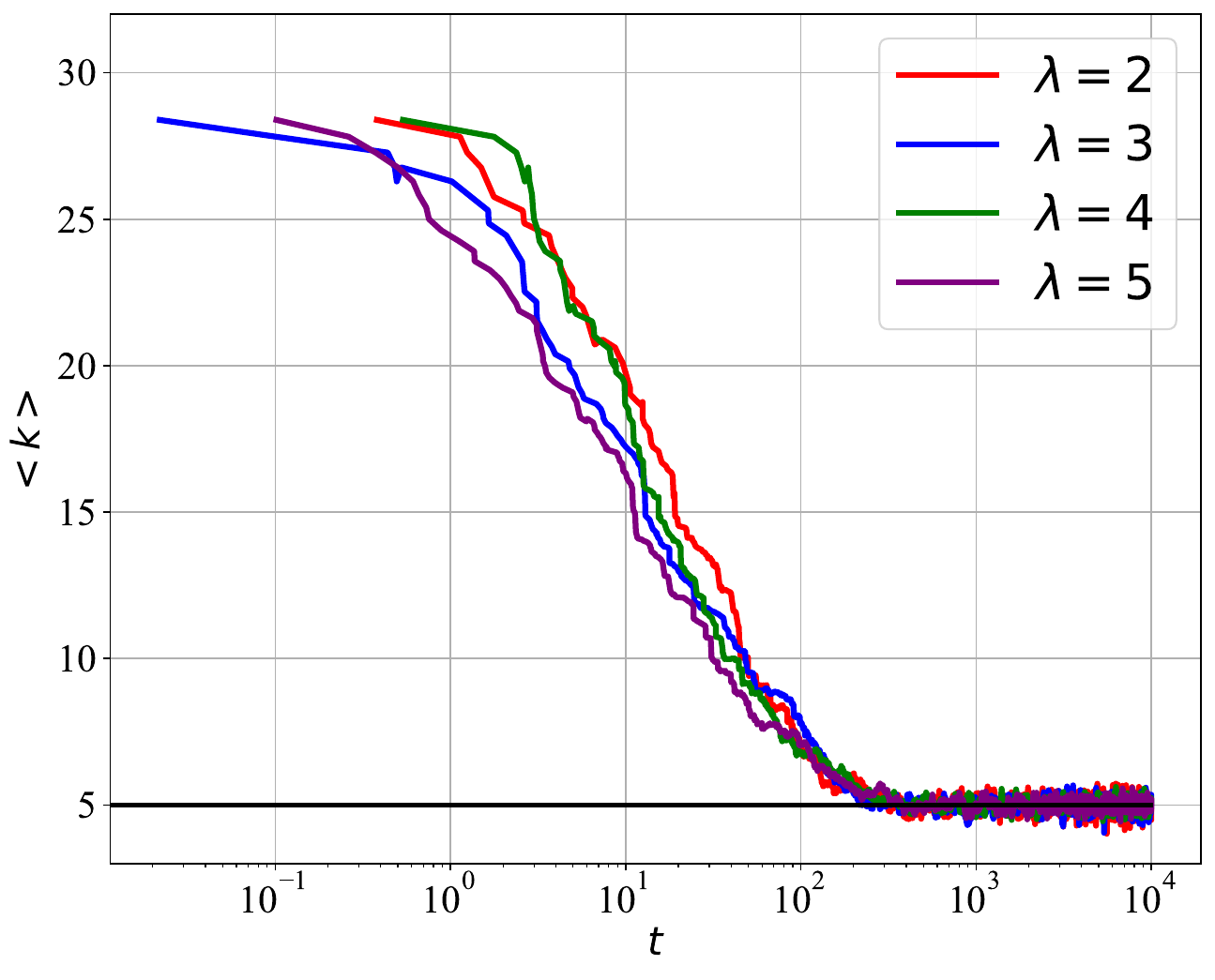}
\label{fig: 5(a)}
}
\subfigure[Evolution with Fixed $\lambda$ (BDEN-uc)]{
\includegraphics[scale=0.25]{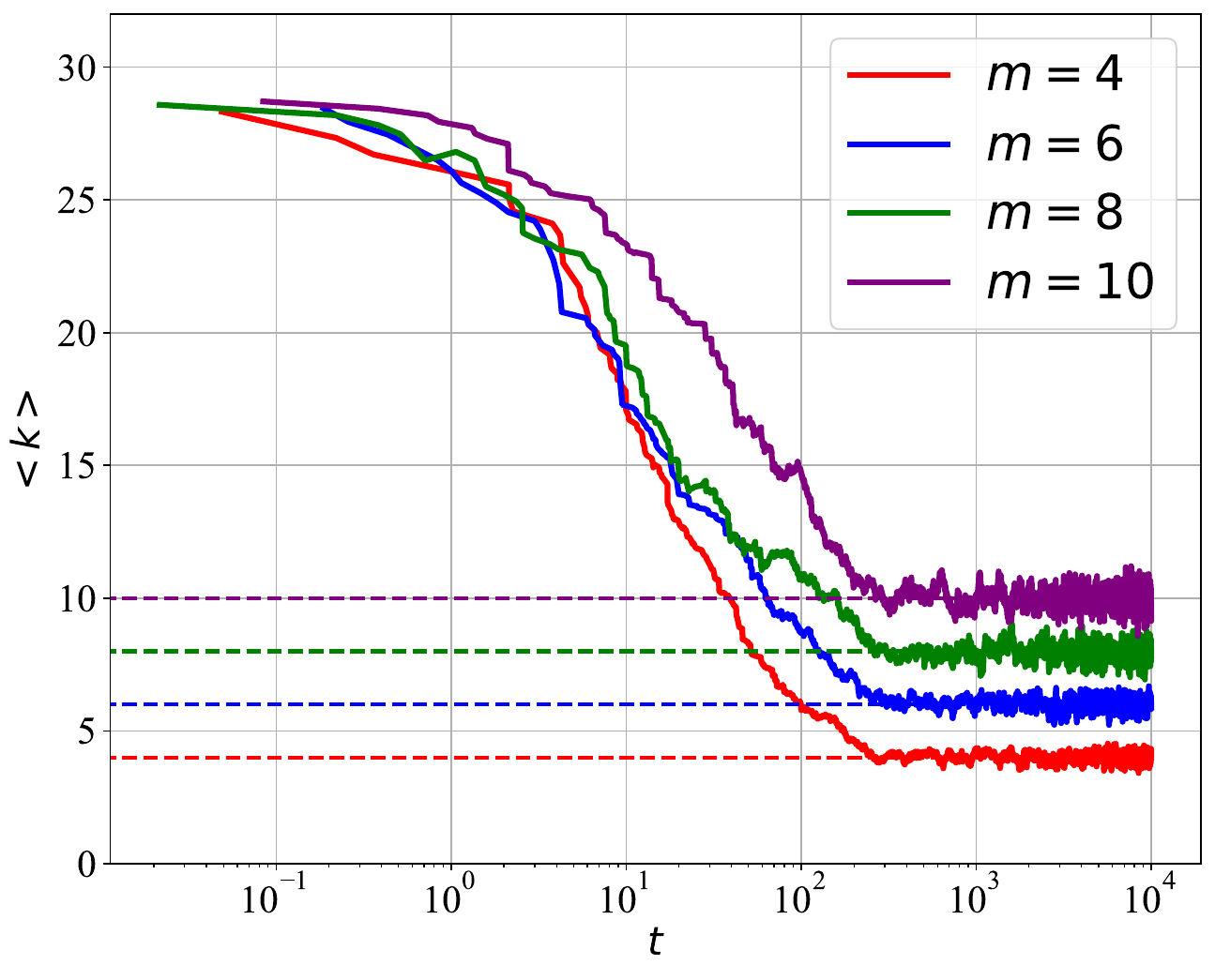}
\label{fig: 5(b)}
}
\subfigure[Mean Degrees Against $\lambda$s (BDEN-uc)]{
\includegraphics[scale=0.25]{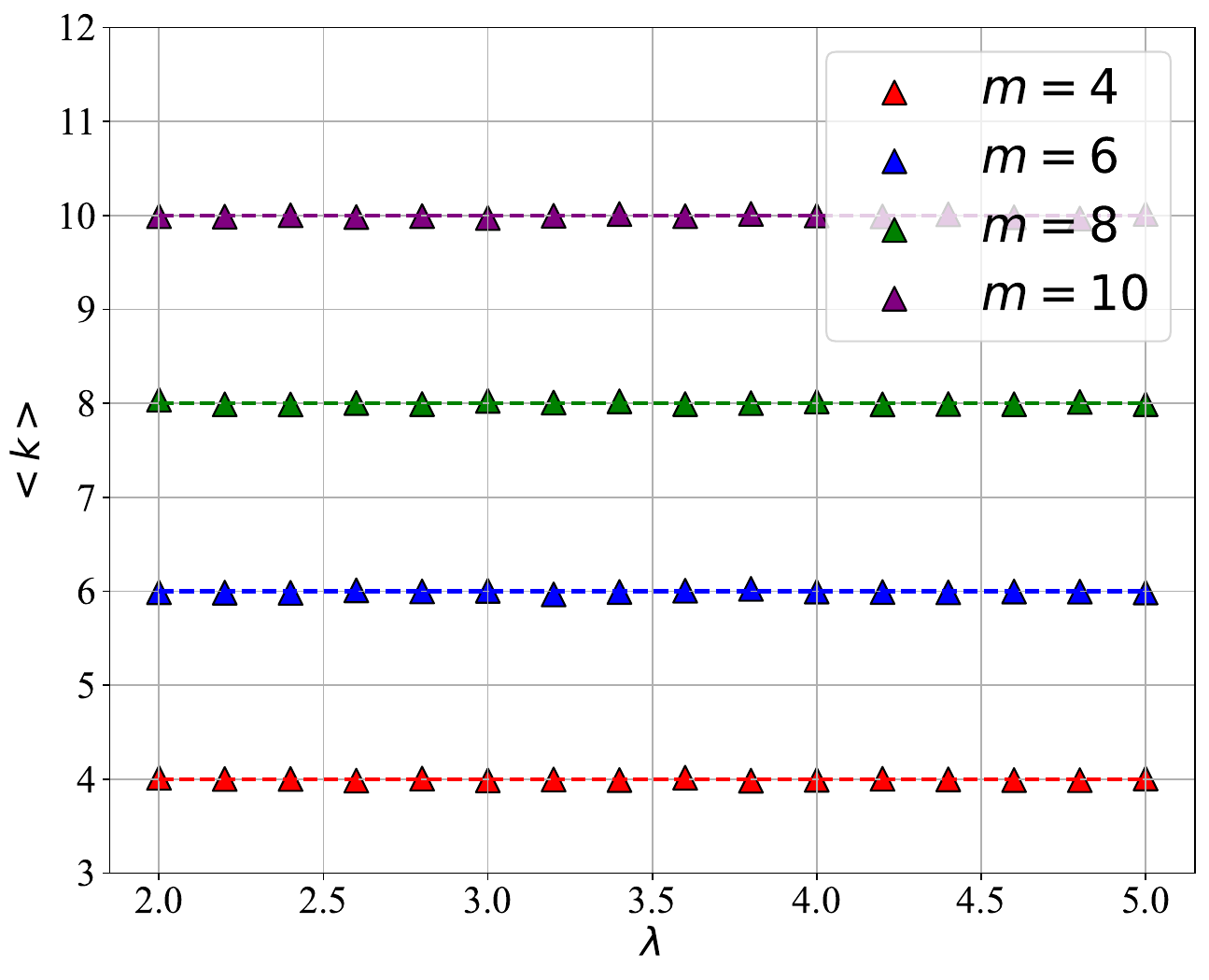}
\label{fig: 5(c)}
}

\subfigure[Evolution with Fixed $m$ (BDEN-pa)]{
\includegraphics[scale=0.25]{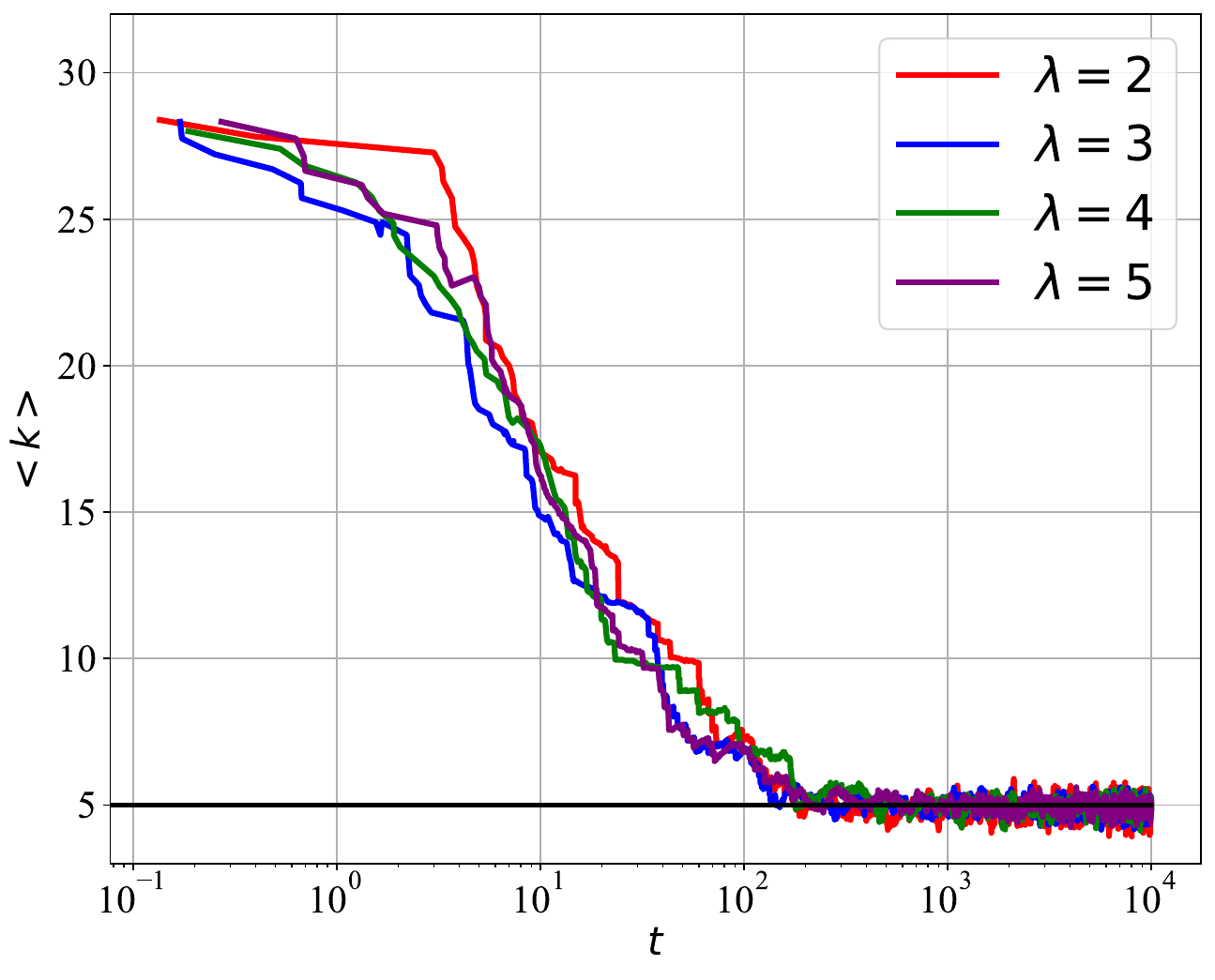}
\label{fig: 5(d)}
}
\subfigure[Evolution with Fixed $\lambda$ (BDEN-pa)]{
\includegraphics[scale=0.25]{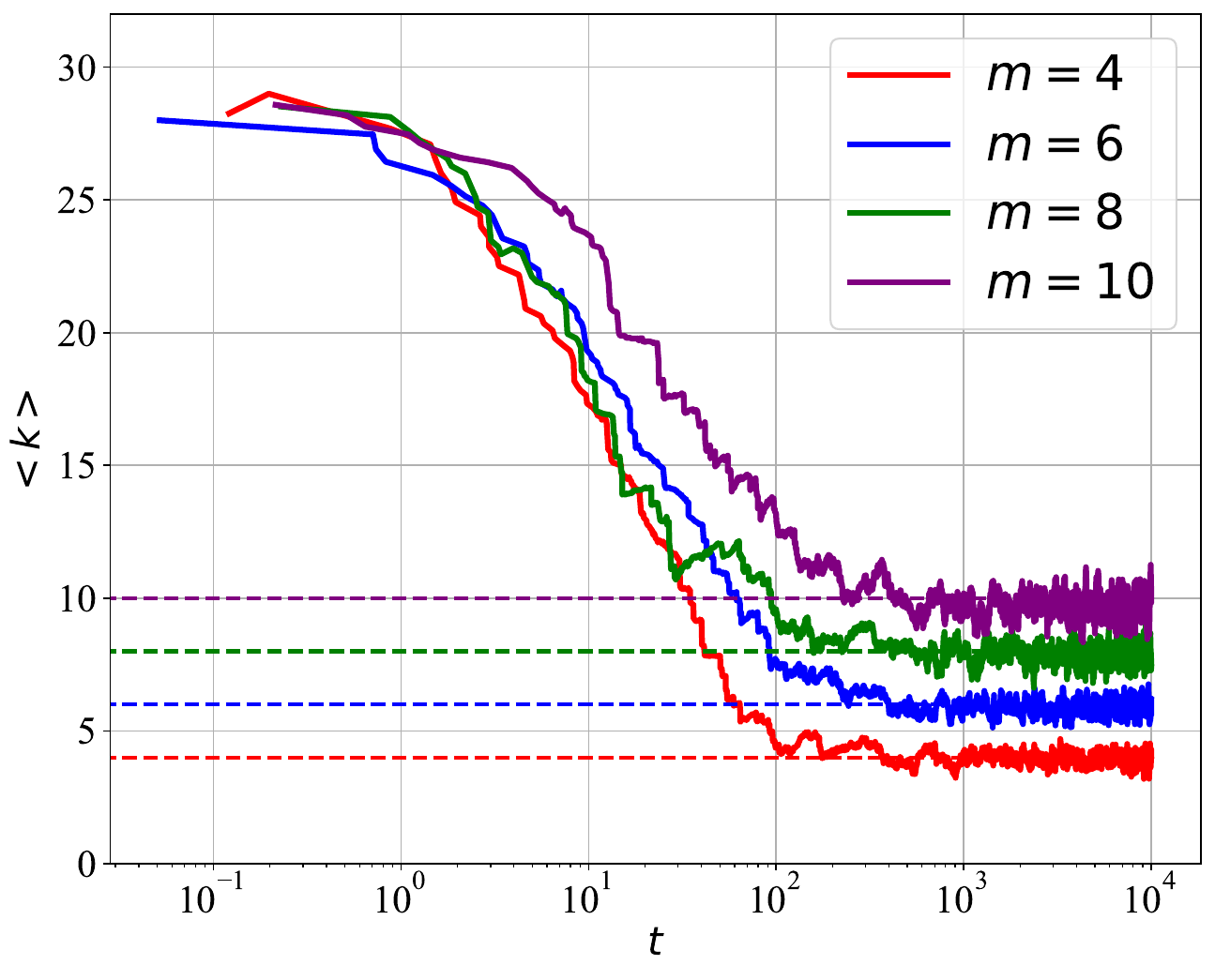}
\label{fig: 5(e)}
}
\subfigure[Mean Degrees Against $\lambda$s (BDEN-pa)]{
\includegraphics[scale=0.25]{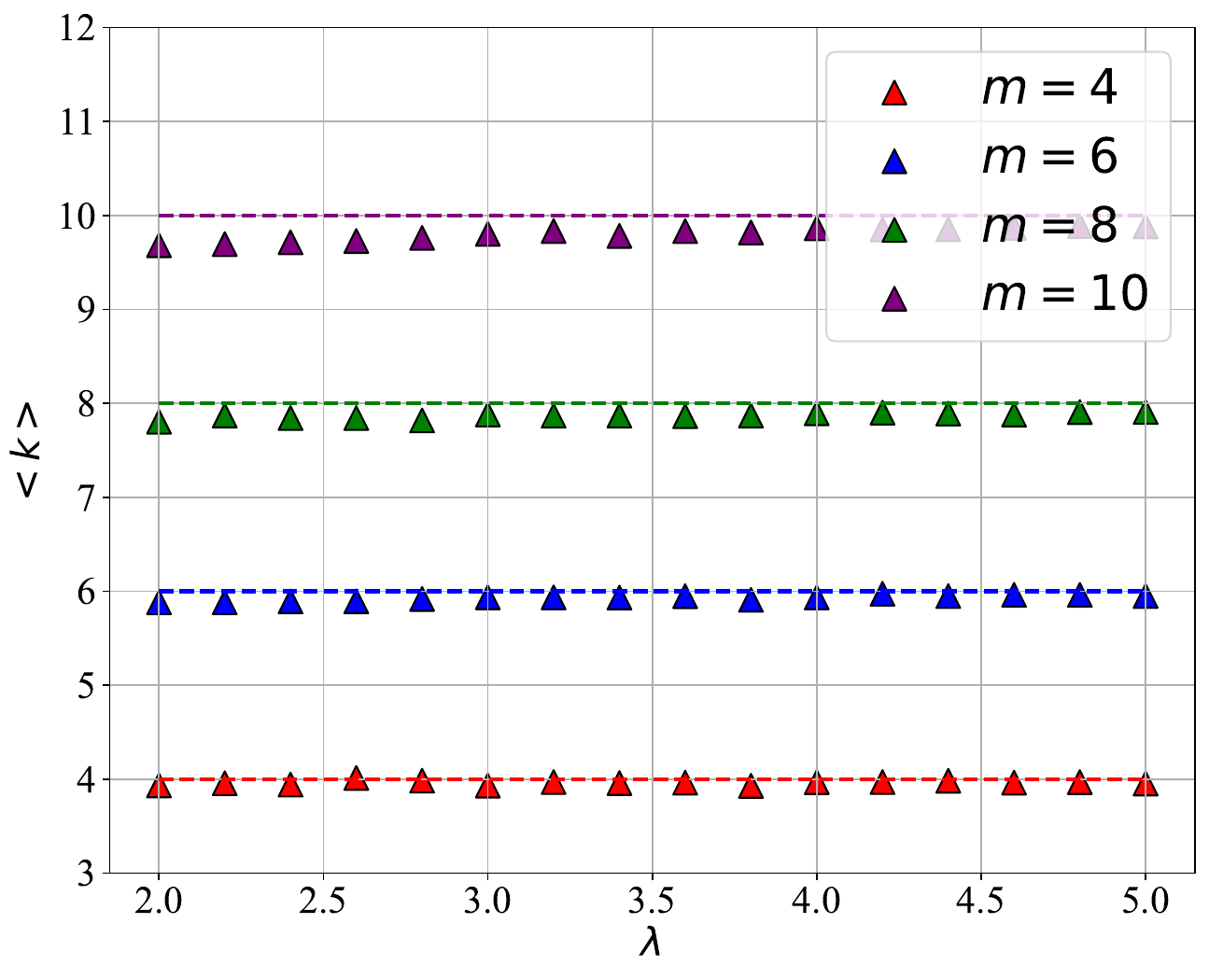}
\label{fig: 5(f)}
}
\caption{\textbf{Results on mean degrees. }(a) The evolution process of mean degrees with the fixed new connection number $m$. (b) The evolution process of mean degrees with the fixed input rate $\lambda$. (c) The curves of mean degrees against $\lambda$s. (a), (b), and (c) are for BDEN-uc, and (d), (e), (f) are for BDEN-pa with the same other settings. In this figure, we set the leave rate of each individual as $\mu=0.01$ and the initial network size $N(0)=30$. In (a) and (d), we fix $m=5$. The $x$-axis is set as time $t$ from $t=0$ to $t=10^4$, and the $y$-axis is set as the mean degree $<k>$. We observe the evolution process of $<k>$s with $\lambda$s$=[2,3,4,5]$. The black line is the theoretical expected degree of each evolving networks. In (b) and (e), we fix $\lambda=3$ and set $m$s$=[4,6,8,10]$. The $x$-axis and $y$-axis are defined the same as (a) and (d). Dashed lines present the theoretical expectation. In (c) and (f), we obtain the average mean degree from $t=10^3$ to $t=10^4$ for $m$s$=[4,6,8,10]$. The $x$-axis is set as the input rate $\lambda\in[2,5]$, where the simulations are performed with the interval $0.2$. The $y$-axis is defined the same as above. (Color online)}
\label{fig:5}
\end{figure*}

Then, we focus on the mean degree of BDEN-uc. Initially, there are $N(0)=30$ individuals in the complete graph. Therefore, the initial mean degree is $\left \langle k \right \rangle=29$. In Fig. \ref{fig: 5(a)}, we primarily present the evolution process of mean degrees from time $t=0$ to $10^4$ with the fixed $m=5$ and $\lambda$s$=[2,3,4,5]$. It is shown that the mean degrees decrease at the beginning of the evolution and become stable around $m=5$ (black line) regardless of the $\lambda$s. Additionally, the fluctuation degree is small, and we can conclude that the mean degrees of all four evolving networks in the performed simulations converge to $5$. Besides, we reckon that the evolution of mean degrees can be considered as stable when $t>10^3$. 

As suggested in Thm. \ref{thm: 2}, the mean degree of the BDEN-uc is irrelevant to both $\lambda$ and $\mu$ but only decided by $m$. Accordingly, in Fig. \ref{fig: 5(b)}, we fix the birth rate $\lambda=3$, $\mu=0.01$ and observe the evolution of mean degrees with $m$s$=[4,6,8,10]$. The dashed lines in Fig. \ref{fig: 5(b)} present the theoretical expectation of the mean degrees. In each group of parameters, the mean degrees decrease with time $t$, and it can be considered that the process has entered a stable state after $t>10^3$. Comparing the stable states and the expectations, we find that the mean degrees oscillate around expectations given by Thm. \ref{thm: 2}. Additionally, the oscillating range is small with a small $m$, resulting in a relatively stable deviation for the evolving network's mean degree of BDEN-uc. In Fig. \ref{fig: 5(c)}, we further present the mean degrees of the evolving networks with different $\lambda$s and $m$s. Each triangular data point is obtained by averaging the mean degrees from $t=10^3$ to $10^4$. Dashed lines are the theoretical expectations of the network mean degrees. Obviously, each curve can be regarded as a primary function with a slope of 0. Therefore, we can employ the Thm. \ref{thm: 2} to describe the mean degree of the BDEN-uc. 

In Figs. \ref{fig: 5(d)}-\ref{fig: 5(f)}, we present the mean degree results for BDEN-pa. As suggested in Figs. \ref{fig: 5(d)} and \ref{fig: 5(e)}, the mean degree of BDEN-pa convergences to the stationary state with the evolution process, which shows similar patterns as we discussed for Figs. \ref{fig: 5(a)} and \ref{fig: 5(b)}. Additionally, the mean degree of BDEN-pa does not show any association with $\lambda$ and $\mu$, but only determined by the new connection number $m$ as the simulation suggests. However, in Fig. \ref{fig: 5(f)}, the statistical expectation of mean degrees are often a little smaller than $m$, which is different from our conclusion in Fig. \ref{fig: 5(c)}. Besides, with the increment of $m$, this gap becomes greater. 

\subsection{Information Diffusion Dynamics}
\begin{figure}                                              
\centering
\subfigure[$\alpha=0.05$]{
\includegraphics[scale=0.17]{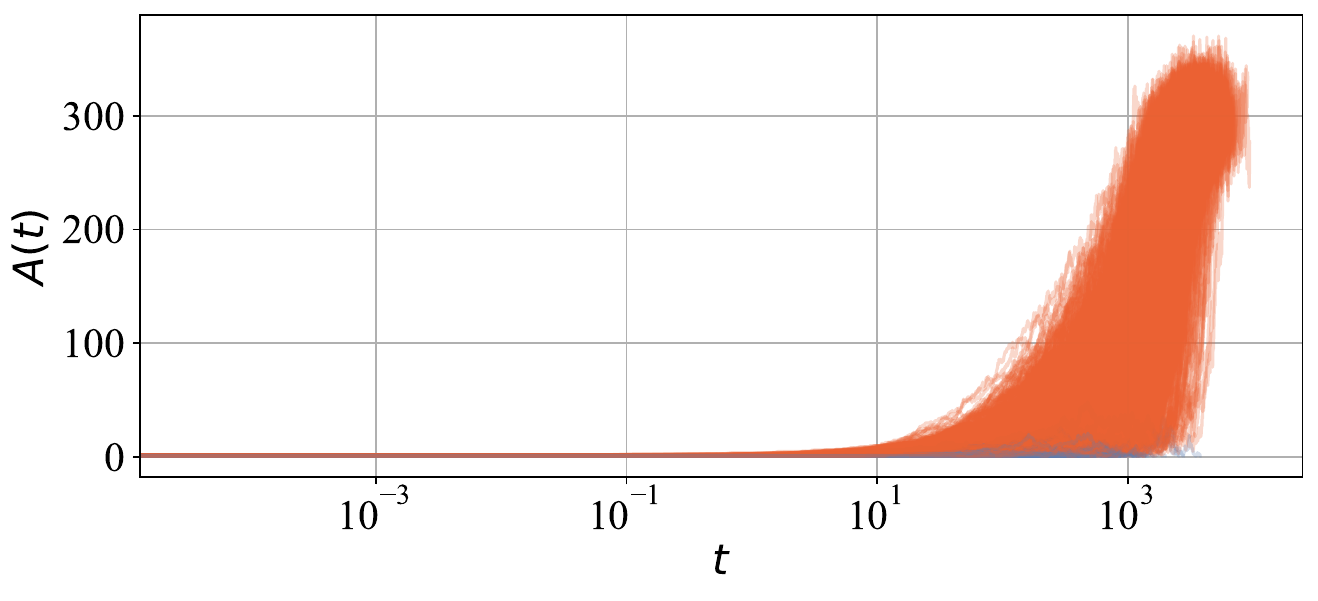}
\label{fig: 6(a)}
}
\subfigure[$\alpha=0.25$]{
\includegraphics[scale=0.17]{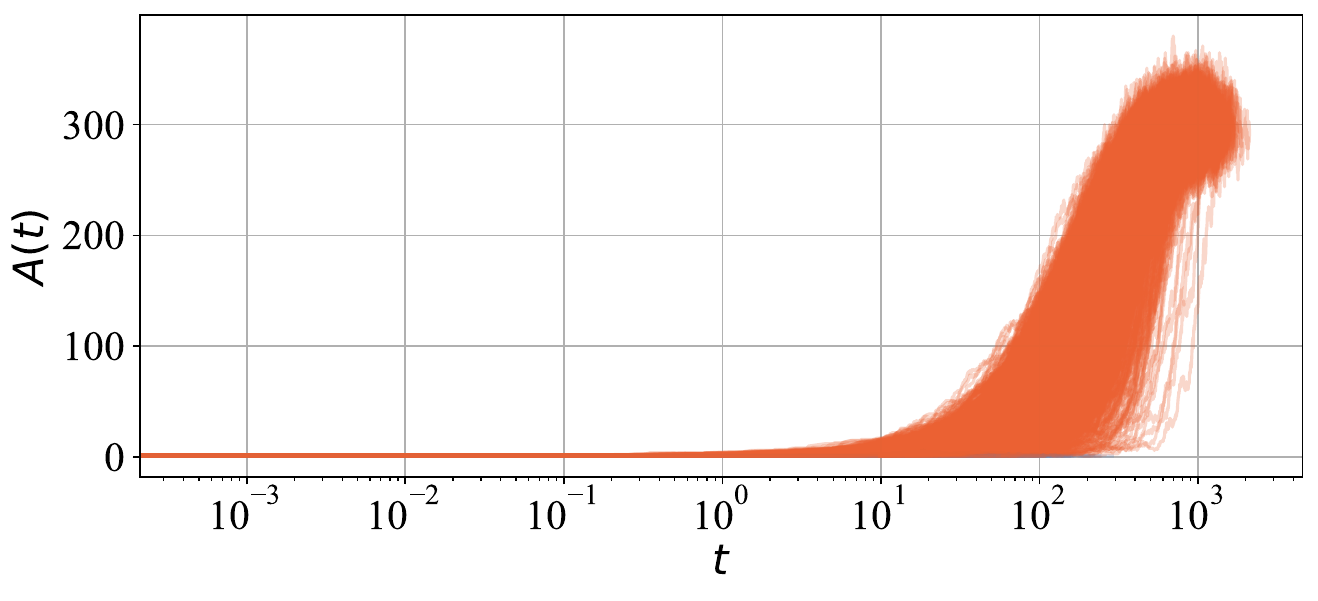}
\label{fig: 6(b)}
}
\caption{\textbf{Evolution of the random drift for BDEN-uc and BDEN-pa. }(a) $\alpha=0.05$. (b) $\alpha=0.25$. In this figure, we set $\lambda=3$ and $m=4$ to show an evolution process of $A(t)$ for the proposed random drift mechanism. The orange and blue curves present the evolution processes that converge to the pure aware and unaware states separately. Note that for each evolution process, we stop collecting the number of aware individuals once $N(t)=A(t)$ or $N(t)=U(t)$, i.e., there is only one state among the population. The initial network is set as a complete graph with $N(0)=30$ vertices and $A(0)=1$ information invader. For each $\alpha$, we perform $1500$ independent and repeated experiments and obtain the above evolution results for both BDEN-uc and BDEN-pa.  (Color online)}
\label{fig:6}
\end{figure}
\begin{figure*}                                              
\centering
\subfigure[BDEN-uc, $m=4$]{
\includegraphics[scale=0.25]{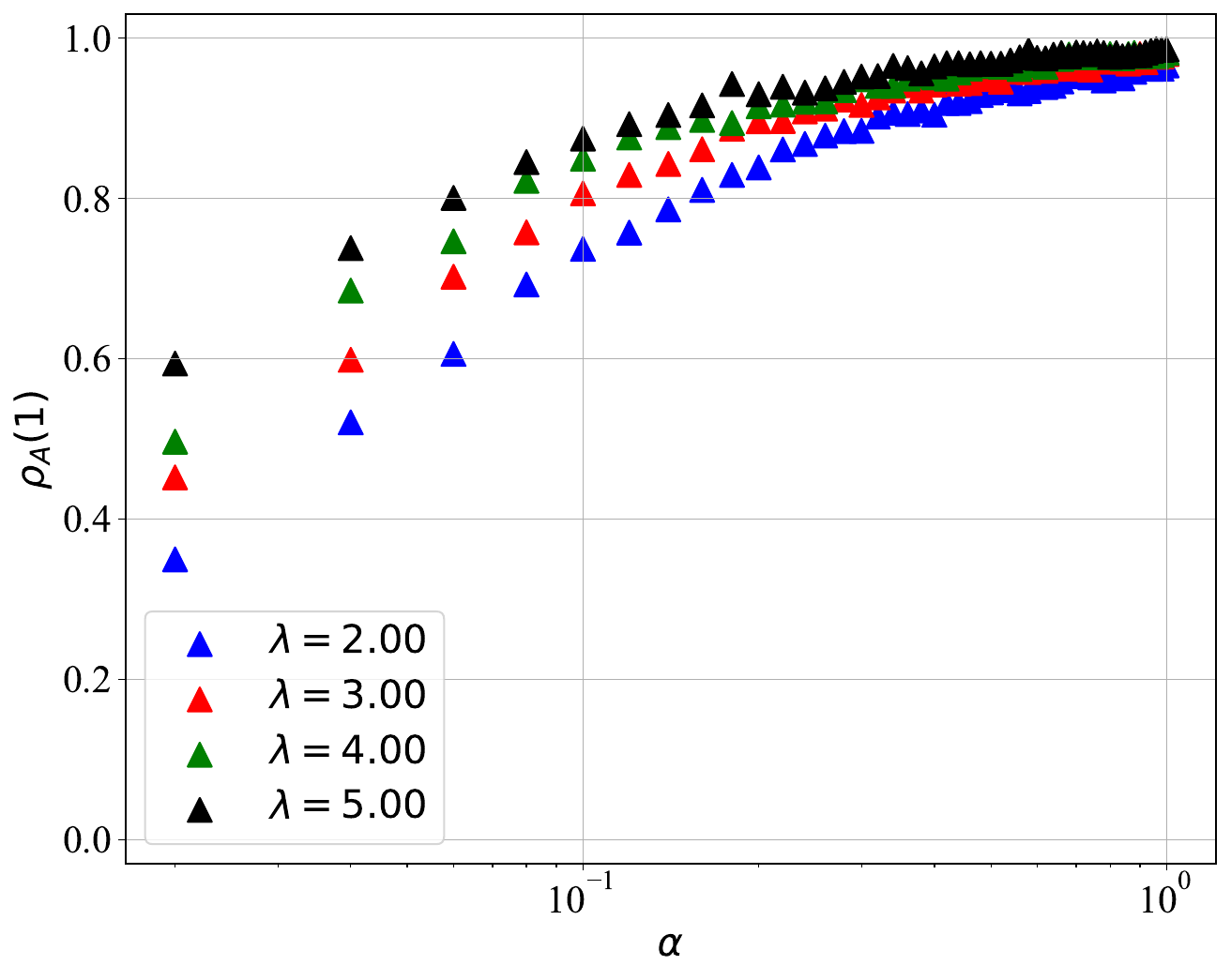}
\label{fig: 7(a)}
}
\subfigure[BDEN-uc, $m=6$]{
\includegraphics[scale=0.25]{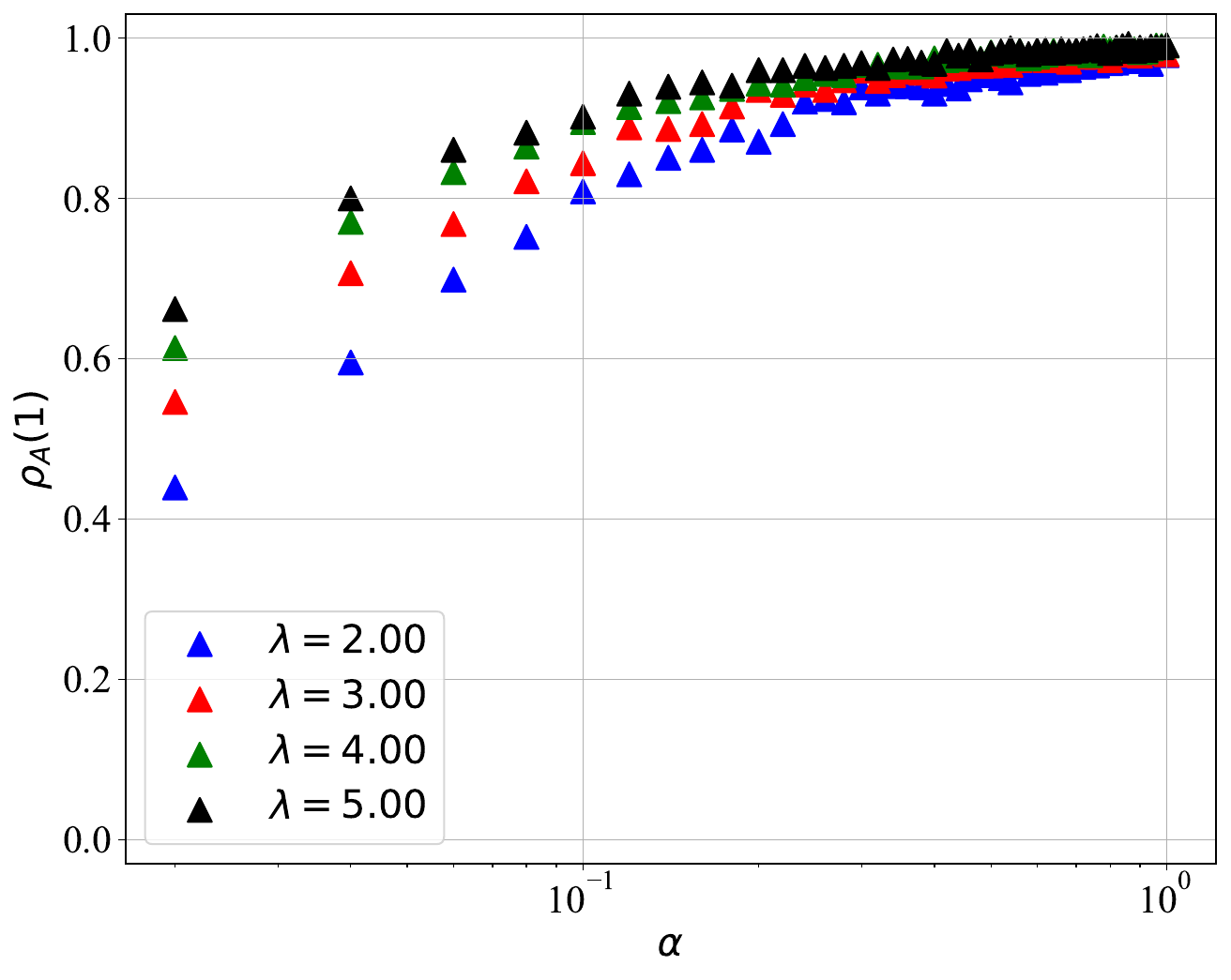}
\label{fig: 7(b)}
}
\subfigure[BDEN-uc, $m=8$]{
\includegraphics[scale=0.25]{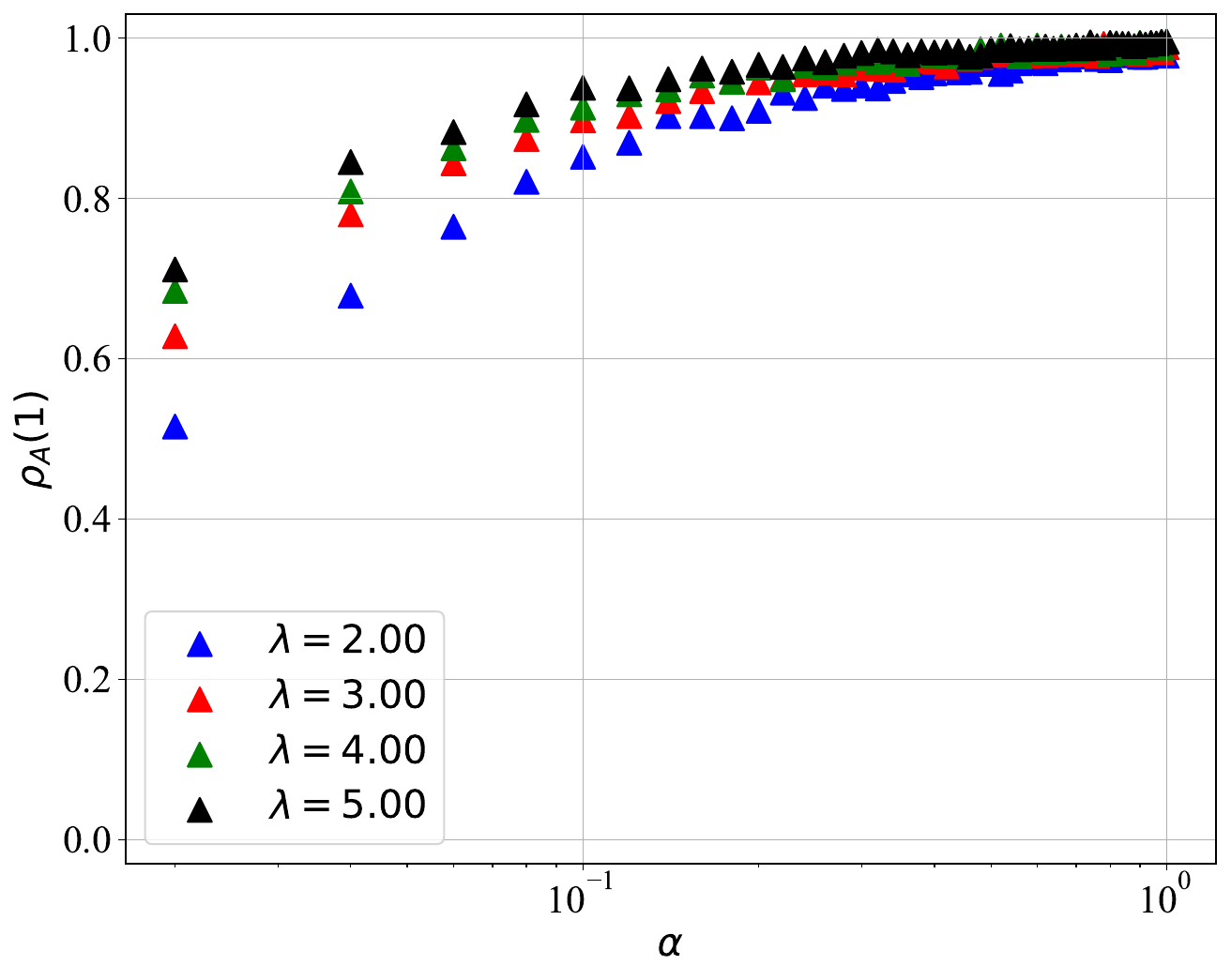}
\label{fig: 7(c)}
}

\subfigure[BDEN-pa, $m=4$]{
\includegraphics[scale=0.25]{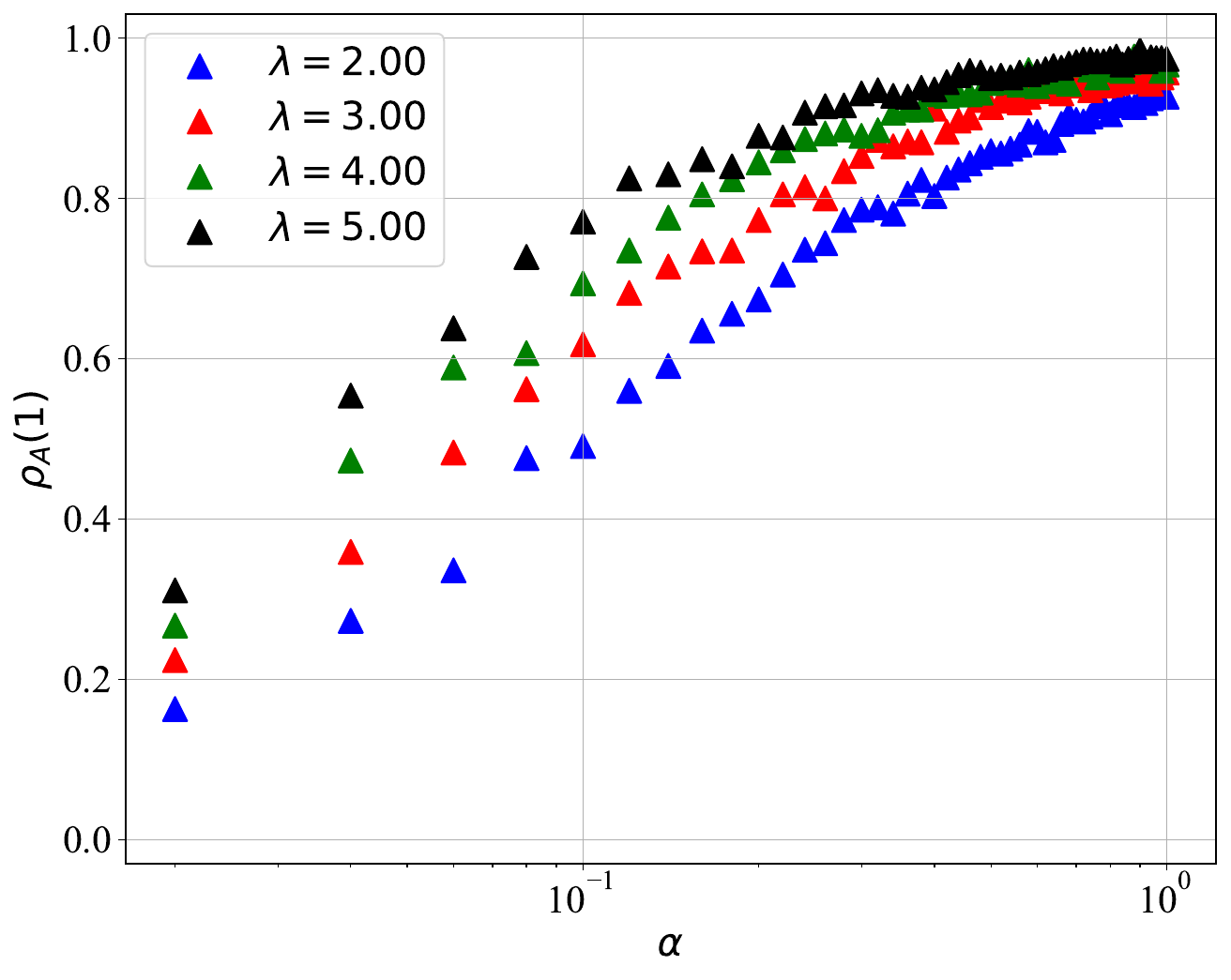}
\label{fig: 7(d)}
}
\subfigure[BDEN-pa, $m=6$]{
\includegraphics[scale=0.25]{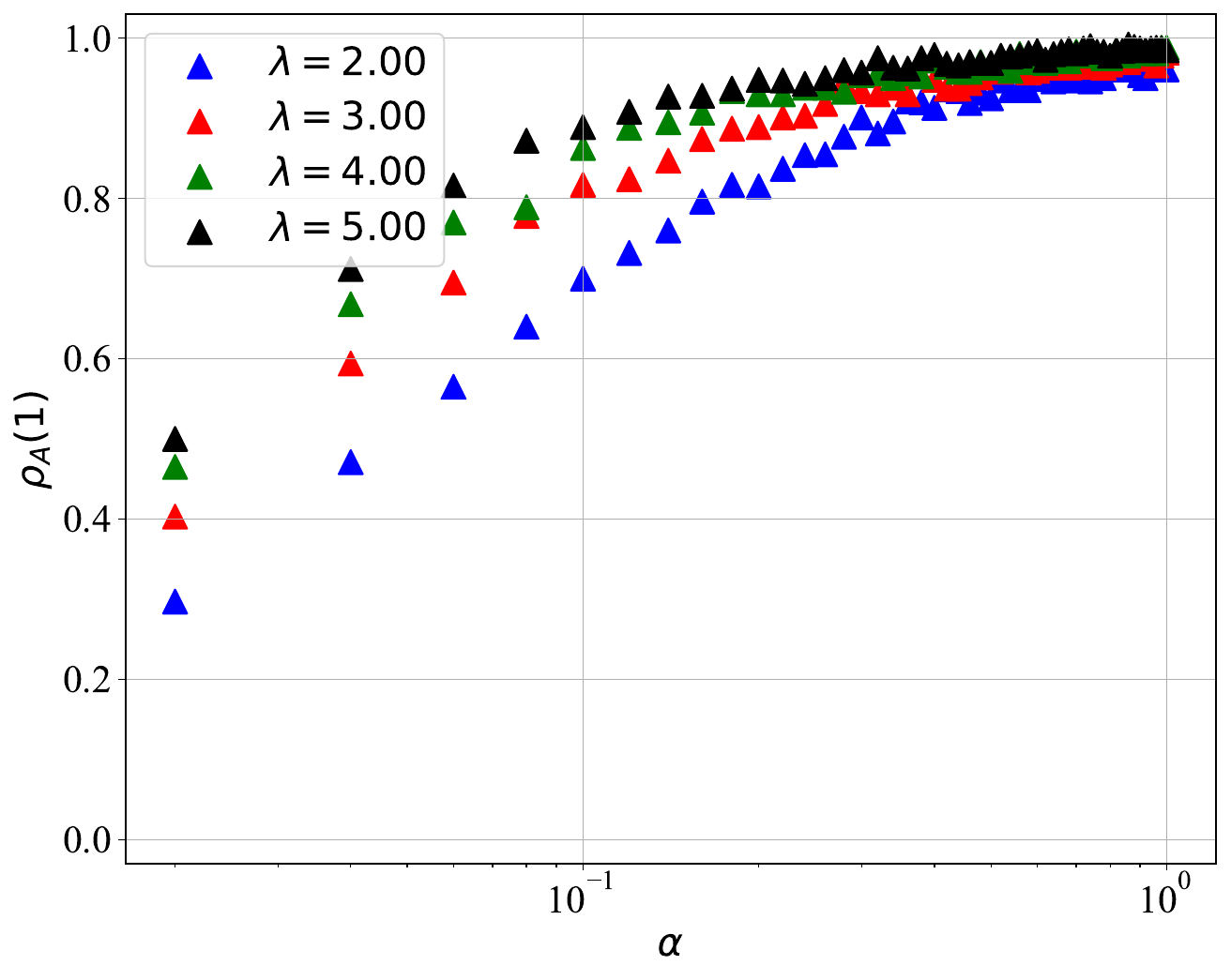}
\label{fig: 7(e)}
}
\subfigure[BDEN-pa, $m=8$]{
\includegraphics[scale=0.25]{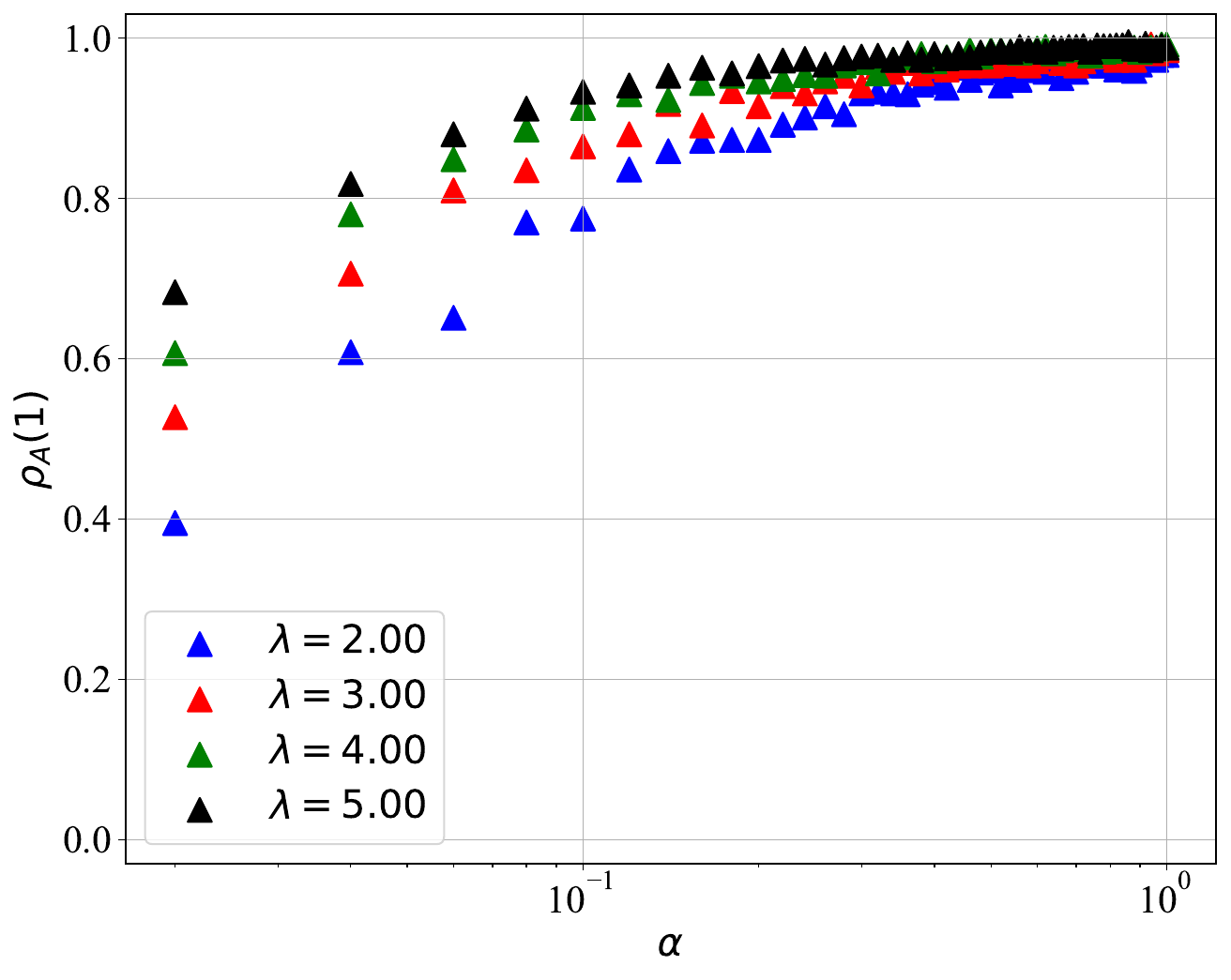}
\label{fig: 7(f)}
}
\caption{\textbf{Fixation probability of the random drift process in the BDEN-uc and BDEN-pa. }(a) BDEN-uc, $m=4$. (b) BDEN-uc, $m=6$. (c) BDEN-uc, $m=8$. (d) BDEN-pa, $m=4$. (e) BDEN-pa, $m=6$. (f) BDEN-pa, $m=8$. This figure presents the fixation probability ($\rho_A(1)$) of the aware individuals with only one information invader that is aware initially for both BDEN-uc and BDEN-pa. We set $\lambda$s$=2,3,4,5$ and $m$s$=4,6,8$ for cross simulation. Additionally, the transmission rate of the information is in the range $[0,1]$. Each fixation probability is obtained by calculating the frequency at which the population finally evolves to the pure aware state in $1500$ experiments. The blue, red, green, and black triangles indicate the results on $\lambda=2,3,4$ and $5$ respectively. (Color online)}
\label{fig:7}
\end{figure*}
Based on Thm. \ref{thm: 3}, we conclude that the population will converge to the pure information state regardless of which diffusion dynamical process it undergoes. Therefore, in the following simulations, we focus on the frequency at which the population evolves to a certain pure state for both BDEN-uc and BDEN-pa. This frequency is also interpreted as the fixation probability $\rho(x)$, where there are $x$ individuals in random positions aware of a certain information at $t=0$ initially. Specifically, we focus on the case $x=1$, meaning that only one individual is aware of this certain information but others are not. This individual can be regarded as an information invader of this population. 
\subsubsection{Random Drift}
For the random drift process, we mainly focus on the fixation probability $\rho_A(1)$ of a specific information where there is only one invading individual aware of the information initially. Primarily, we present the evolution process of the random drift mechanism in Fig. \ref{fig:6} to confirm the Thm. \ref{thm: 3}. Figs. \ref{fig: 6(a)} and \ref{fig: 6(b)} show the evolution of aware individuals with time, where we set $\alpha=0.05,0.25$ and perform $1500$ independent and repeated simulations for both BDEN-uc and BDEN-pa respectively. For a better presentation, we stop plotting the evolution process as the population information reaches the pure state. We notice that if the time is large enough, all curves come to breakpoints, i.e., all evolutionary processes reach the pure states under both BDEN-uc and BDEN-pa. Accordingly, we can conclude that Thm. \ref{thm: 3} for the random drift process is confirmed. 

Based on this conclusion, we then explore the influence of different parameters on the fixation probabilities. In Fig. \ref{fig:7}, we show the fixation probability with only one information invader for the random drift process initially with $\lambda$s$=[2,3,4,5]$ and $m$s$=[4,6,8]$ for BDEN-uc. Each fixation probability is obtained by over $1500$ independent and repeated simulations. Obviously, it is more possible for information to occupy the whole population with a high transmission rate $\alpha$. Besides, all fixation probabilities almost converge to one if the transmission rate $\alpha$ is large enough. Then, let us consider the effect of different birth rates $\lambda$s on the fixation probabilities. For all results on Figs. \ref{fig: 7(a)}, \ref{fig: 7(b)}, and \ref{fig: 7(c)}, the fixation probability is often high if the population has a high birth rate with the same transmission rate $\alpha$. For example, in Fig. \ref{fig: 7(a)}, the fixation probabilities from numerical simulations are $0.737$, $0.806$, $0.850$, and $0.875$ for $\lambda$s$=[2,3,4,5]$ respectively if $\alpha=0.10$. Moreover, if the birth rate $\lambda$ and the transmission rate $\alpha$ are fixed, the fixation probability of the information increases with a new connection number of the population. As an illustration, if we set $\alpha=0.16$ and $\lambda=3$, the fixation probabilities for $m$s$=[4,6,8]$ are $0.861$, $0.893$, and $0.934$ respectively. In addition, the effect of both $\lambda$s and $m$s on the fixation probabilities are weakened and even almost eliminated with the increase of the transmission rates $\alpha$s. That is to say, although it is shown that the transmission rate and new connection number are higher the better for information to spread widely among the BDEN-uc, information with a relatively high transmission rate almost certainly takes root in the population regardless of the birth rate $\lambda$ and the number of new links $m$. 

\begin{figure}                                              
\centering
\subfigure[$b/c=5$]{
\includegraphics[scale=0.17]{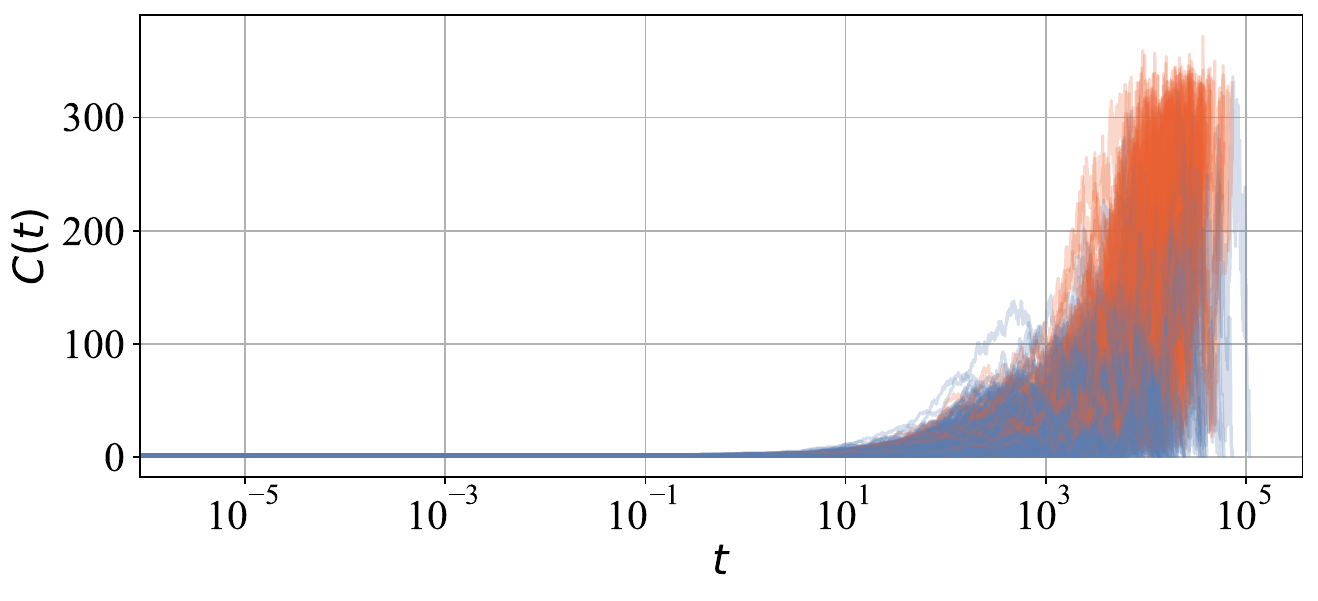}
\label{fig: 8(a)}
}
\subfigure[$b/c=15$]{
\includegraphics[scale=0.17]{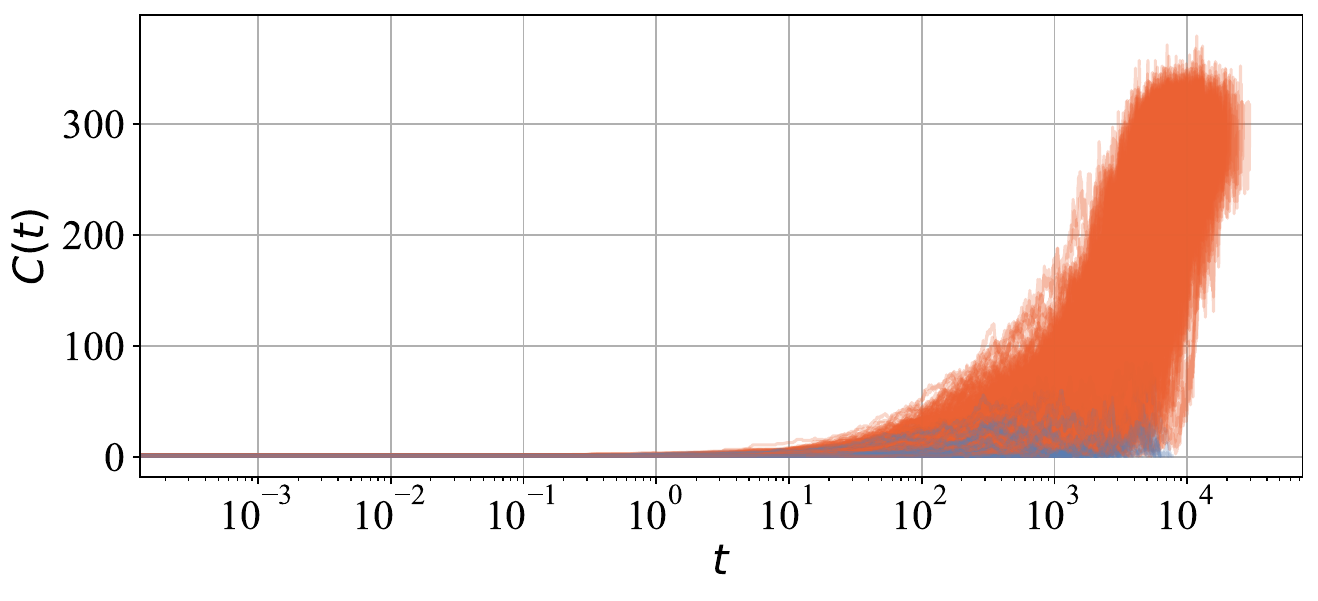}
\label{fig: 8(b)}
}
\caption{\textbf{Evolution of the natural selection for BDEN-uc and BDEN-pa. }(a) $b/c=5$. (b) $b/c=15$. In this figure, we set $\lambda=3$, $m=4$, and $c=1$ to show an evolution process of $C(t)$ for the proposed natural selection mechanism. The orange and blue curves present the evolution processes that converge to the pure aware state of $C$ and the pure aware state of $D$ separately. Note that for each evolution process, we stop collecting the number of aware individuals once $N(t)=C(t)$ or $N(t)=D(t)$. The initial network is set as a complete graph with $N(0)=30$ vertices and $C(0)=1$ information invader. For each $\alpha$, we perform $1500$ independent and repeated experiments and obtain the above evolution results for both BDEN-uc and BDEN-pa. (Color online)}
\label{fig: 8}
\end{figure}

For BDEN-pa, the trend of the fixation probability with the transmission rate $\alpha$ is similar to BDEN-uc. However, BDEN-pa holds a lower fixation probability given the same couple of parameters, especially when $\alpha<0.4$. As the increment of the transmission rate $\alpha$, the differences are getting smaller. Therefore, we conclude that the preferential attachment mechanism can inhibit the fixation of the information in the birth-death evolving population comparing to uniform connection. 
\subsubsection{Natural Selection}
As stated previously, there are two kinds of information $C$ and $D$ conflicting in the natural selection process. Therefore, we focus on the fixation probability $\rho_C(1)$ of the information $C$ for both BDEN-uc and BDEN-pa. To simplify the issue, we rewrite the payoff matrix $M$ into the form of the prisoner's dilemma, which is a common model for conflict between two sides in society. In detail, we have $R=b-c$, $S=-c$, $T=b$, and $P=0$ where $R>P$. In prisoner's dilemmas, an individual's local interests often conflict with overall interests. Therefore, the egoistic information $D$ can be influenced and occupied by other individuals' altruistic information $C$ to ensure global fitness. For convenience, we call $b/c$ the cost-to-benefit ratio of the information diffusion process. 

Similar to the above contents, we first verify Thm. \ref{thm: 3} for natural selection by showing the evolution process of the population in Fig. \ref{fig: 8}. With fixed $\lambda=3$, $m=4$, and $c=1$, we set $b=5$ and $15$ for cross simulation respectively in Figs. \ref{fig: 8(a)} and \ref{fig: 8(b)}. Each figure contains $1500$ independent evolution processes for both BDEN-uc and BDEN-pa that evolve to the pure $C$ or $D$ states colored by orange and blue respectively. Besides, we stop plotting the evolution processes if the population reaches the information absorption state. As Fig. \ref{fig: 8} suggests, all curves converge to the pure $C$ or $D$ states. Therefore, the information dynamics in the population only converge to $A(t)=0$ or $D(t)=0$ as Thm. \ref{thm: 3} for natural selection presents. 

\begin{figure*}                                              
\centering
\subfigure[BDEN-uc, $m=4$]{
\includegraphics[scale=0.25]{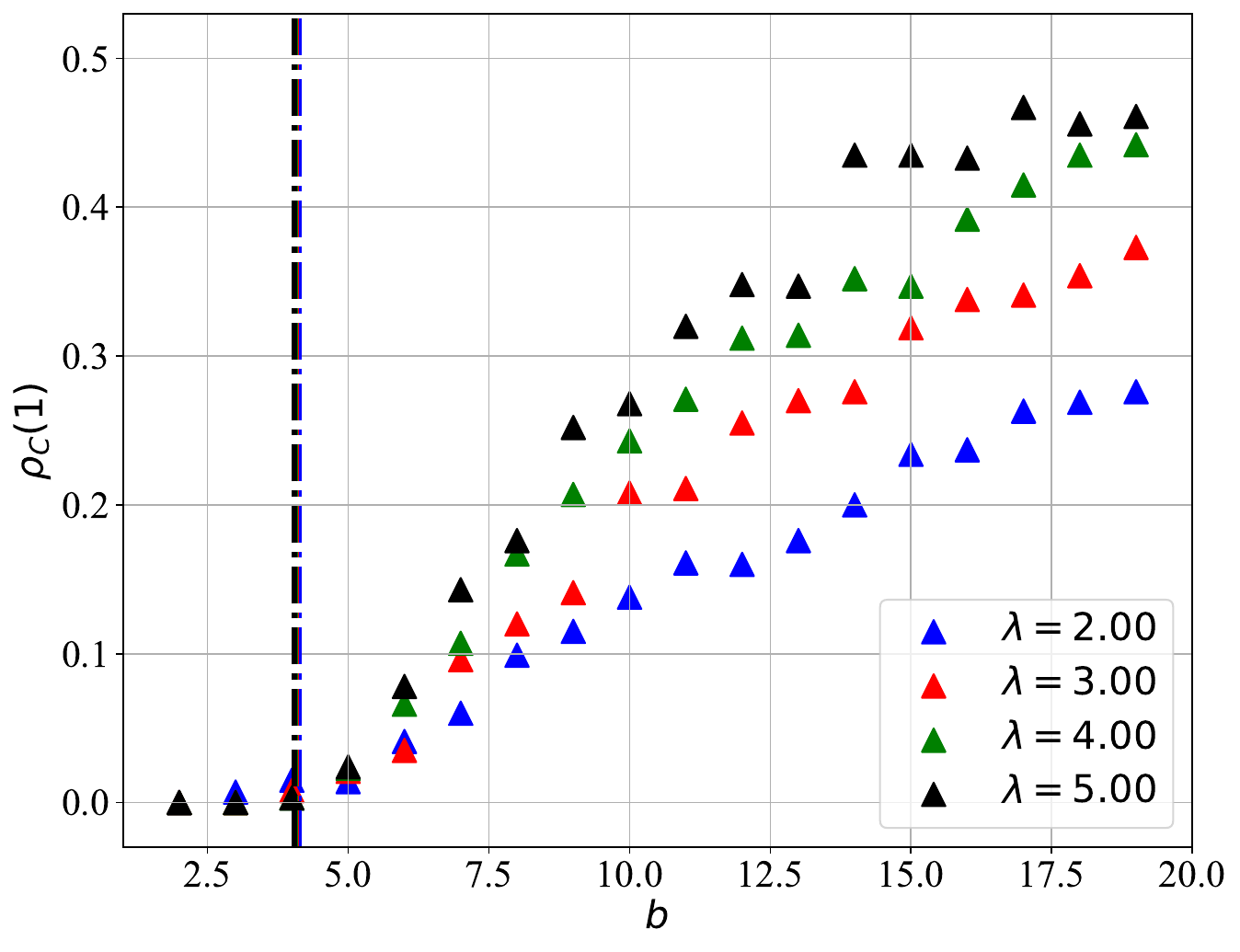}
\label{fig: 9(a)}
}
\subfigure[BDEN-uc, $m=6$]{
\includegraphics[scale=0.25]{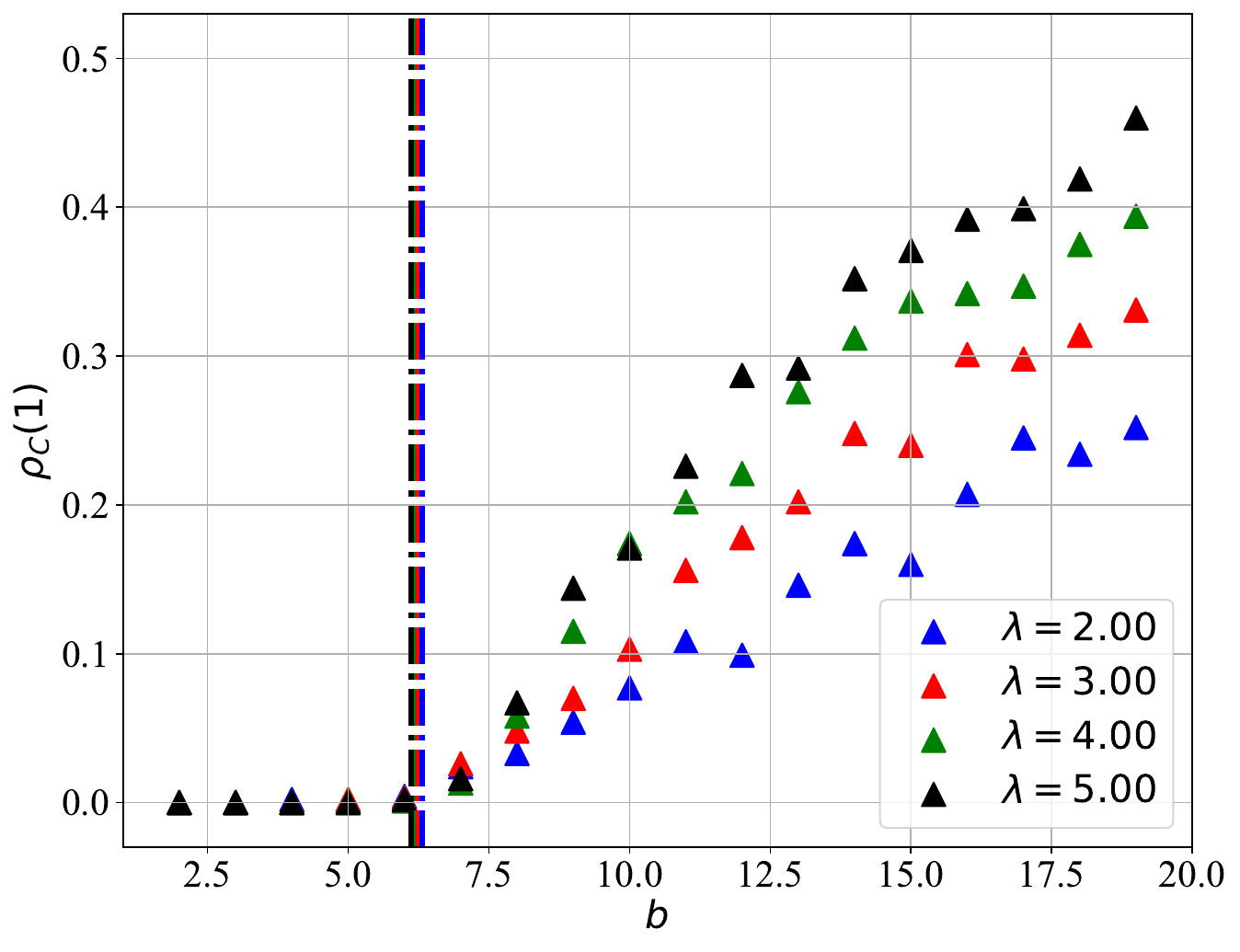}
\label{fig: 9(b)}
}
\subfigure[BDEN-uc, $m=8$]{
\includegraphics[scale=0.25]{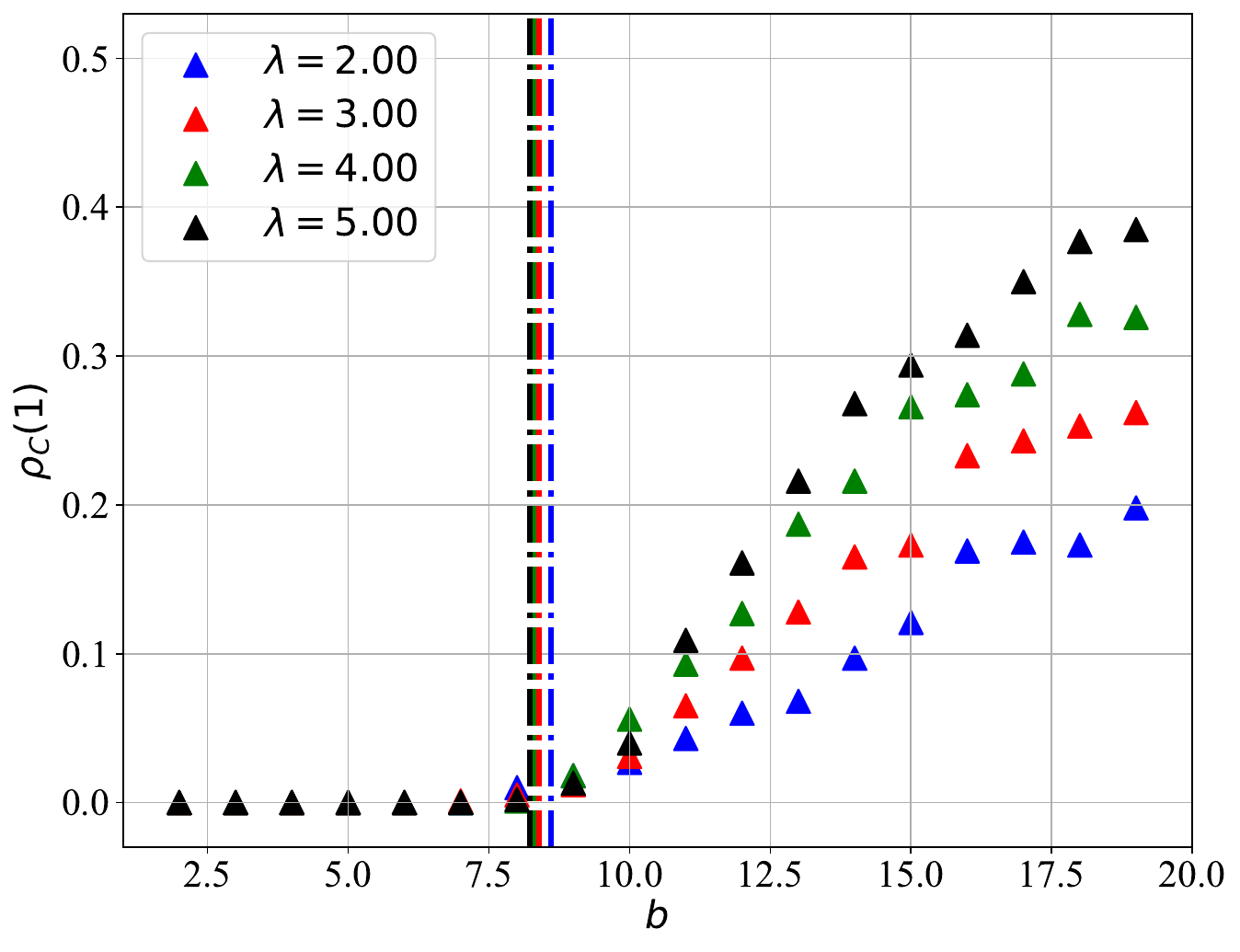}
\label{fig: 9(c)}
}

\subfigure[BDEN-pa, $m=4$]{
\includegraphics[scale=0.25]{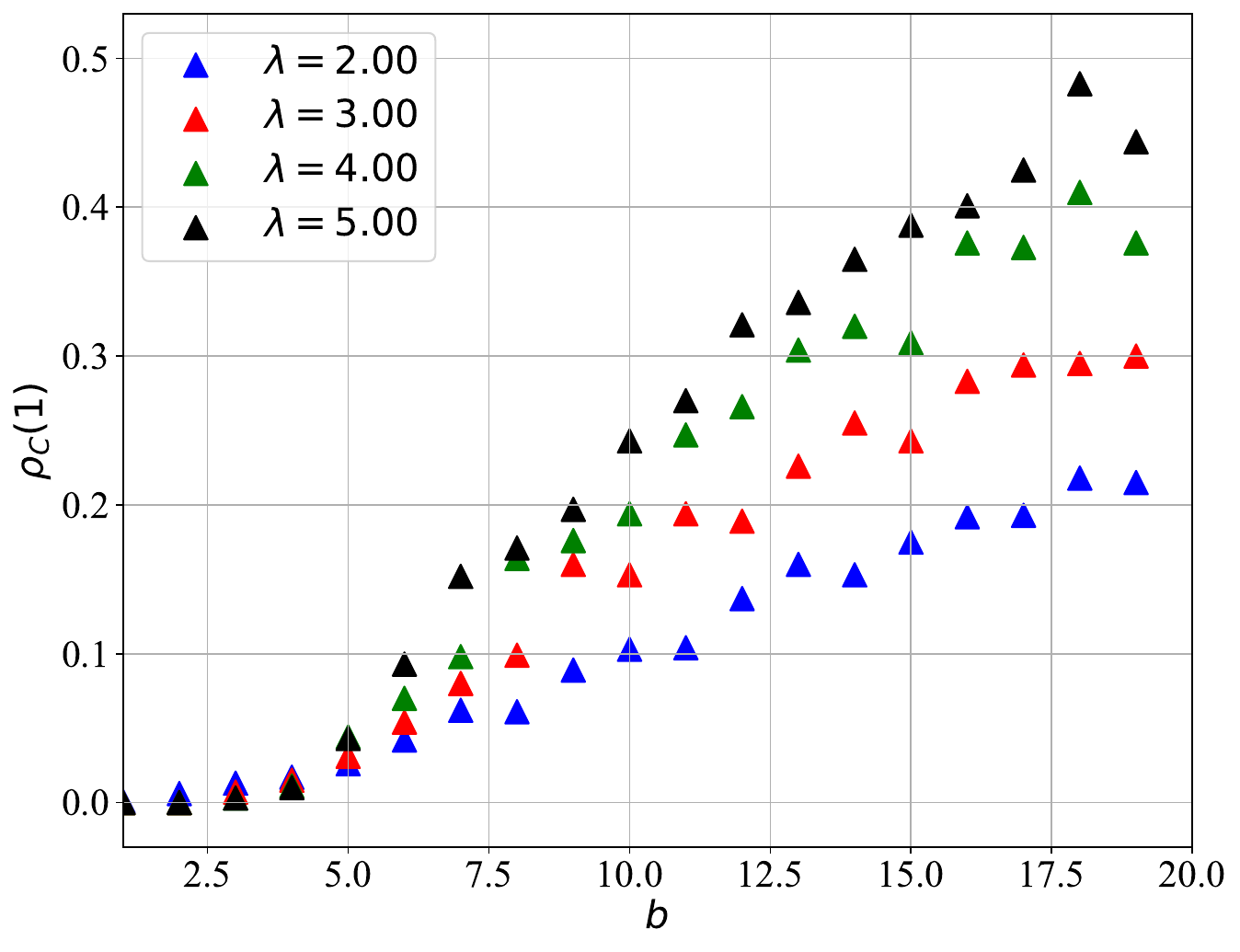}
\label{fig: 9(d)}
}
\subfigure[BDEN-pa, $m=6$]{
\includegraphics[scale=0.25]{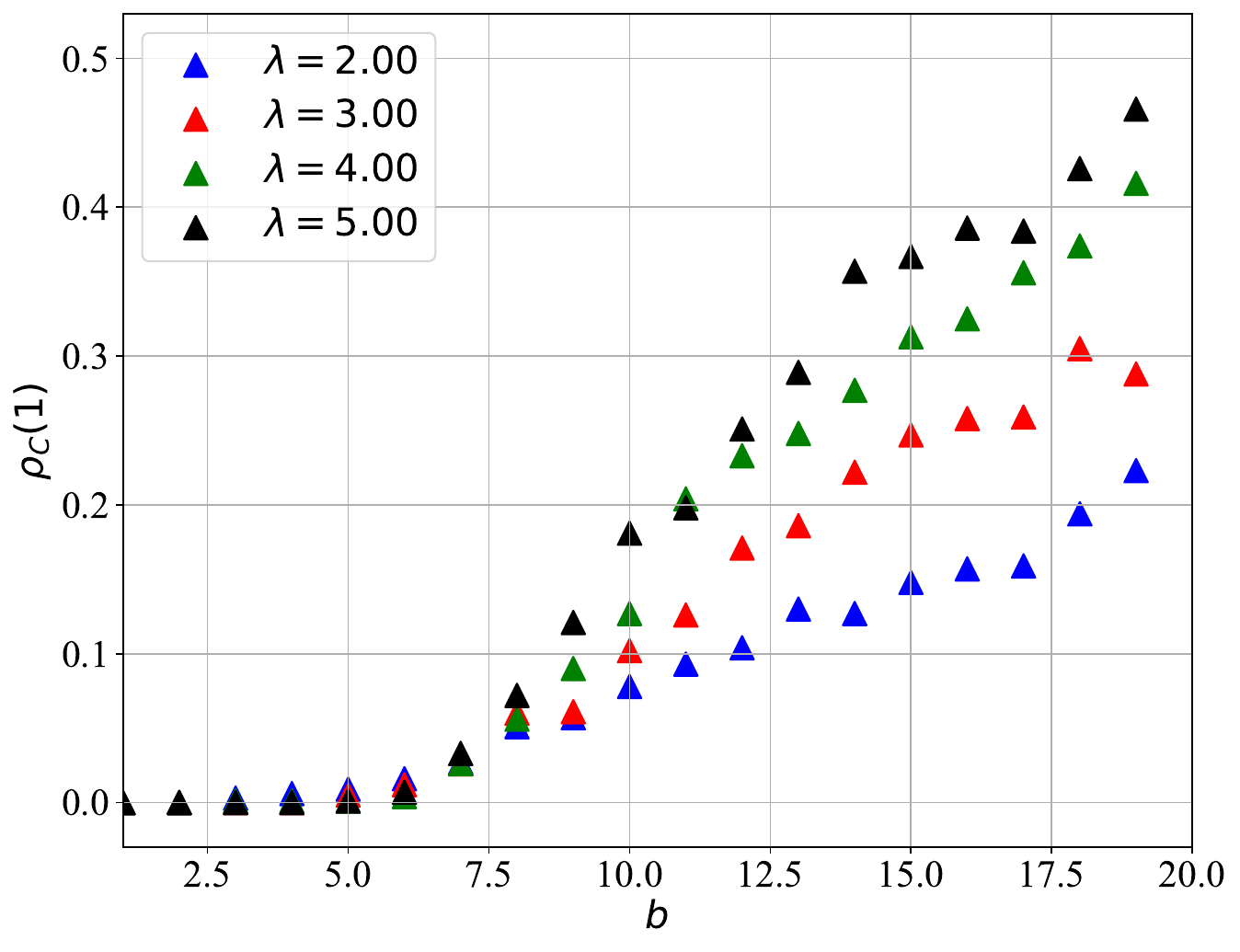}
\label{fig: 9(e)}
}
\subfigure[BDEN-pa, $m=8$]{
\includegraphics[scale=0.25]{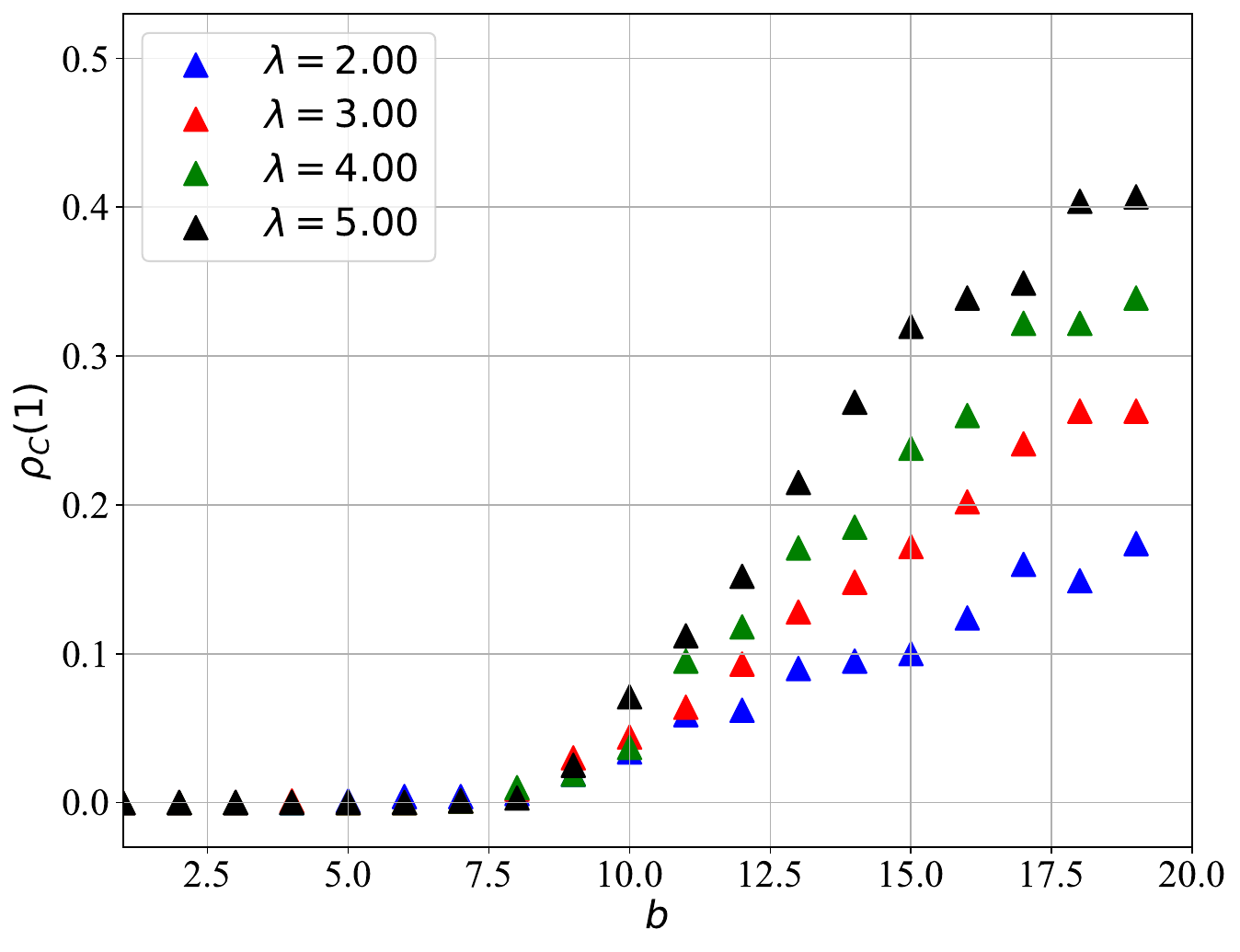}
\label{fig: 9(f)}
}
\caption{\textbf{Fixation probability of the natural selection process in the BDEN-uc and BDEN-pa. }(a) BDEN-uc, $m=4$. (b) BDEN-uc, $m=6$. (c) BDEN-uc, $m=8$. (d) BDEN-pa, $m=4$. (e) BDEN-pa, $m=6$. (f) BDEN-pa, $m=8$. This figure presents the fixation probability ($\rho_C(1)$) of $C$ individuals with only one invading $C$ player initially. Theoretical solutions are shown in vertical lines. We set $\lambda$s$=2,3,4,5$ and $m$s$=4,6,8$ for cross simulation. We perform the simulation on $c=1$ and $b\in[2,19]$ with the step one. Each fixation probability is obtained by calculating the frequency at which the population finally evolves to the pure state of $C$ in $1500$ experiments. The blue, red, green, and black triangles indicate the results on $\lambda=2,3,4$ and $5$ respectively. (Color online)}
\label{fig: 9}
\end{figure*}

Then, let us consider the fixation probability of the information $C$ that is beneficial to global fitness. In Fig. \ref{fig: 9}, we initialize the network as a complete graph with $N(0)=30$ vertices with only one invading $C$ player while other individuals hold the information $D$ for both BDEN-uc and BDEN-pa. Similar to the results on random drift, we set $\lambda$s$=[2,3,4,5]$ and $m$s$=[4,6,8]$ for cross simulations. The fixation probability is obtained in the range $b\in[2,19]$ with space one, where each point in Fig. \ref{fig: 9} is obtained by calculating the ratio of pure $C$ states in $1500$ independent and repeated experiments. In Figs. \ref{fig: 9(a)}-\ref{fig: 9(c)}, we present the natural selection process for BDEN-uc. With the increase of the cost-to-benefit ratio $b/c$, the fixation probability of the information $C$ increases as well. Besides, a population with a high birth rate $\lambda$ has a large fixation probability of the information $C$, i.e., a high birth rate is beneficial for the global fitness of the population. For example, if $m=4$ and $b=10$ (equivalent to the cost-to-benefit ratio $b/c=10$ here), the fixation probabilities of the information $C$ are $0.138$, $0.208$, $0.243$, and $0.268$ for $\lambda$s$=[2,3,4,5]$ respectively in Fig. \ref{fig: 9(a)}. Additionally, with the increase of the new connection number $m$, the altruistic information $C$ becomes more likely to take over the population. Note that there are critical cost-to-benefit values for the absorption state of $C$ players. These critical values are around $b=4$, $6$, and $8$ in Figs. \ref{fig: 9(a)}, \ref{fig: 9(b)}, and \ref{fig: 9(c)} respectively, which are approximate to the corresponding new connection numbers $m$s. 

In Figs. \ref{fig: 9(d)}-\ref{fig: 9(f)}, we present the fixation probability results for BDEN-pa. The results show that BDEN-pa shows both similar conclusions and the critical benefit-to-cost ratio (b/c) for the emergence of cooperative behaviors. Therefore, BDEN-uc and BDEN-pa have akin effects on the fixation probability of cooperative behavior. Compared to previous results on random drift, we find that the preferential attachment has a smaller effect on the natural selection process. Note that the difference between these two models lies in the connection mechanism, leading to the fact that the birth-death of network individuals reduces the effect of uniform connection and preferential attachment in the evolution of cooperation. 

Based on Thm. \ref{thm: 4}, we show the condition that $C$ behavior is favored in the network for BDEN-uc. Our simulation results verified this theorem as Figs. \ref{fig: 9(a)}-\ref{fig: 9(c)}, where the theoretical solutions are presented as vertical lines. It is worth noting that this conclusion seems applicable for BDEN-pa as well, probably because of the similar degree properties and the reduction of heterogeneity from leaving vertices. 
\subsection{Natural Selection Starting From Real Networks}
\begin{figure}                                              
\centering
\subfigure[Visual Cortex]{
\includegraphics[scale=0.18]{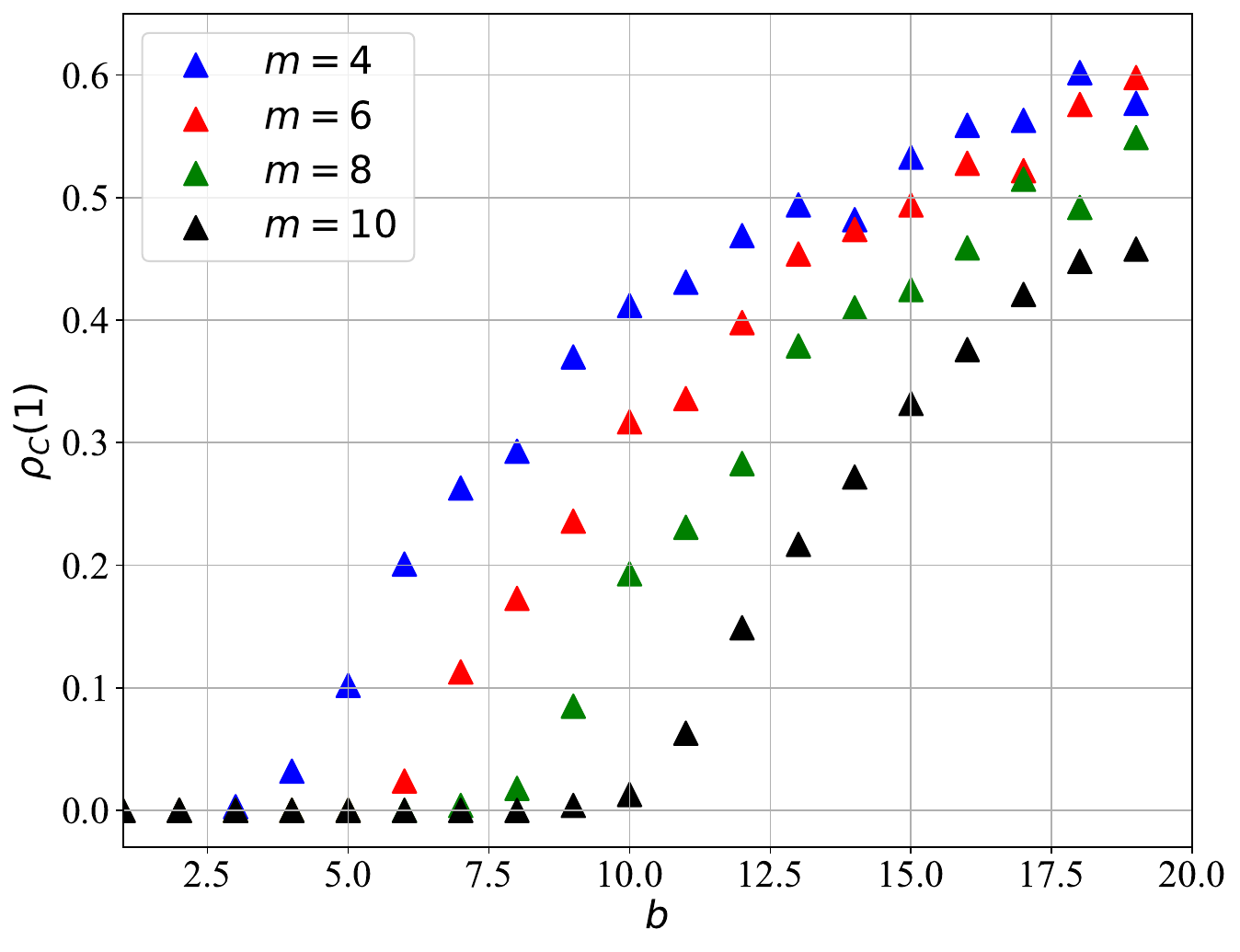}
\label{fig: 10(a)}
}
\subfigure[Gr\'{e}vy's zebras]{
\includegraphics[scale=0.18]{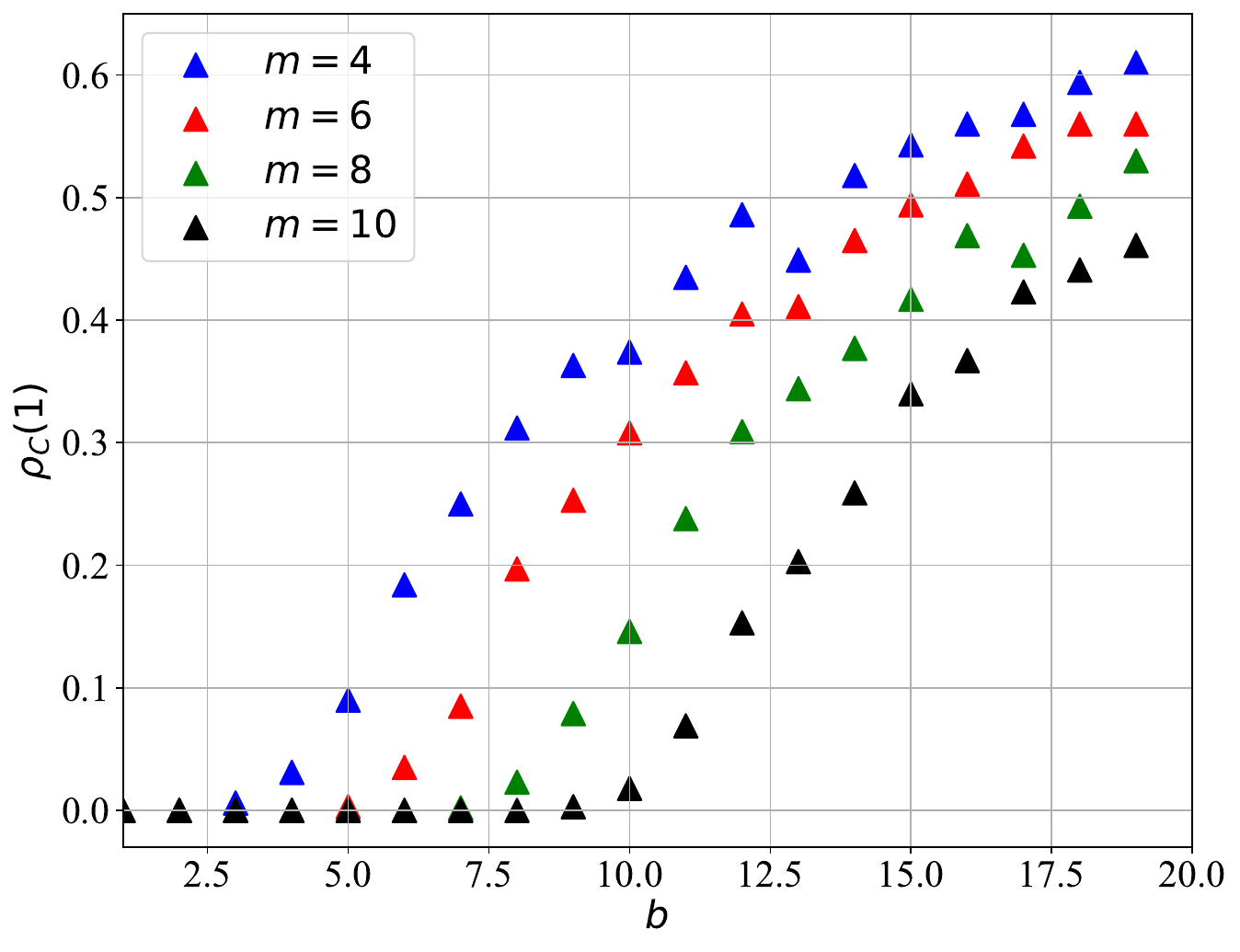}
\label{fig: 10(b)}
}
\caption{\textbf{Fixation probability of the natural selection process starting from real-world data sets. }(a) Visual cortex network. (b) Gr\'{e}vy's zebra network. This figure presents the fixation probability ($\rho_C(1)$) of the altruistic information $C$ with only one invading $C$ player initially starting from real networks. We set $\lambda=5$ and $\mu=0.01$ for the evolving network. The payoff parameters are set as $c=1$ and $b\in[1,20)$. Each fixation probability is obtained by calculating the frequency at which the population finally evolves to the pure state of $C$ in $1500$ experiments. The blue, red, green, and black triangles indicate the results on $m=4,6,8$ and $10$ respectively.  (Color online)}
\label{fig: 10}
\end{figure}
Our proposed evolving network model starts from a random network with new individuals' arrival and old individuals' departure. Owing to the fact that the fitness-driven phase transition is more commonly found in real systems, we now investigate the natural selection in evolving networks that start from real-world network data sets. The employed real networks are the Moreno zebra network \cite{konect}, which describes interactions between 28 Gr\'{e}vy's zebras \cite{konect:2017:moreno_zebra}, \cite{konect:sundaresan}, and the visual cortex network \cite{nr}, which describes fiber tracts that connect one vertex to another between 29 vertices \cite{bigbrain}. 

In Fig. \ref{fig: 10}, we obtain the fixation probability of the $C$ behavior in the evolving birth-death network with parameters $\lambda=5$ and $\mu=0.01$. As shown in Fig. \ref{fig: 10(a)}, altruistic behavior bursts to take root in the population as the cost-to-benefit $(b/c)>m$. If $b/c$ is greater than the new connection number $m$, the fixation probability increases as $b/c$ increases. Similar fixation probability values are obtained in Fig. \ref{fig: 10(b)}, and the same condition for cooperation also holds in Fig. \ref{fig: 9}, demonstrating the fact that the different initial network structures only have little influence on the collective behaviors for natural selection with the evolution of networks and strategies. 
\section{Conclusions and Outlooks}
\label{sec: conclusions and outlooks}
In this paper, we present an information diffusion model based on the birth-death process in an evolving networked population. Regarding the connection mechanisms of new individuals, we introduce both the uniform connection (BDEN-uc) and preferential attachment (BDEN-pa). We analyze the network structure and find that the expected size of the population depends on the birth-death rates, while the network's mean degree is determined by the new connection number. We propose two mechanisms, random drift and natural selection, to describe the propagation of single information and competition between two conflicting pieces of information, respectively. We show that the information in the population evolves to an absorption state where all individuals have the same status, regardless of the proposed mechanism.

To verify our analysis, we conduct simulations and investigate the fixation probability of certain information with only one information invader at the start of evolution. We find that high birth rates, new connection numbers, and transmission rates increase the fixation probability of certain information in the population undergoing random drift. However, increasing the transmission rate eliminates the effects of birth rates and new connection numbers. In the conflict between two pieces of information, high birth rates, and cost-to-benefit ratios promote the emergence of altruistic information, but a higher new connection number is counterproductive. The critical cost-to-benefit ratio of the emergence of altruistic information coincides with the rule proposed in previous research. Our results also suggest that BDEN-uc allows more probability for fixation of the information than the BDEN-pa in random drift (especially with a small transmission rate), while BDEN-uc and BDEN-pa have almost the same effects for the emergence of cooperative behavior in natural selection. 

This work supplements the theoretical tools in describing the properties of the evolving birth-death networks and the framework to study the corresponding dynamic processes. Our primary achievement is the establishment of the diffusion model in the evolving birth-death network model and the conclusions about network topology and dynamic processes. Additionally, we performed a stochastic analysis of the evolving network properties and dynamic process features and confirmed theorems by simulations. However, the model can be further improved by considering non-Poisson processes for the birth-death of individuals, and by incorporating the possibility of existing individuals changing their state continuously. These modifications could increase the universality and strictness of the information diffusion model. Moreover, we consider that the individuals only change their state as they join the population for simplicity of the theoretical analysis. If the existing individuals also change their state at some speed continuously, although this greatly increases the complexity of the model, the universality and strictness may be improved. 
\appendices
\ifCLASSOPTIONcaptionsoff
  \newpage
\fi

\begin{IEEEbiography}
[{\includegraphics[width=1in,height=1.25in,clip,keepaspectratio]{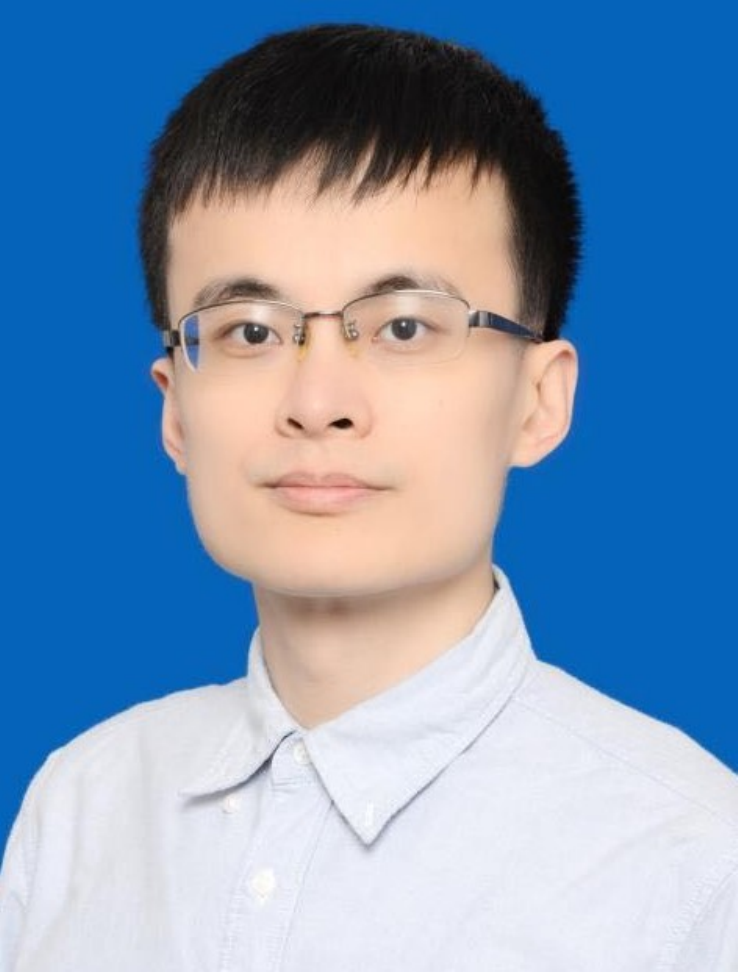}}]{Minyu Feng}
received the B.S. degree in mathematics from the University of Electronic Science and Technology of China in 2010; the Ph.D. degree in computer science from the University of Electronic Science and Technology of China in 2018. From 2016 to 2017, he was a joint-training Ph.D. student with the Potsdam Institute for Climate Impact Research, Germany, and Humboldt University, Berlin, Germany. Since 2019, he has been an associate professor at the College of Artificial Intelligence, Southwest University, Chongqing, China. 

Dr. Feng is a Senior Member of IEEE, China Computer Federation (CCF) and Chinese Association of Automation (CAA). He currently serves as an editorial board member of PLOS ONE, PLOS Complex Systems, Frontiers in Physics and International Journal of Mathematics for Industry. 

Dr. Feng's research interests include complex networks, complex systems, stochastic processes, evolutionary games theory and digital epidemiology.
\end{IEEEbiography}
\begin{IEEEbiography}[{\includegraphics[width=1in,height=1.25in,clip,keepaspectratio]{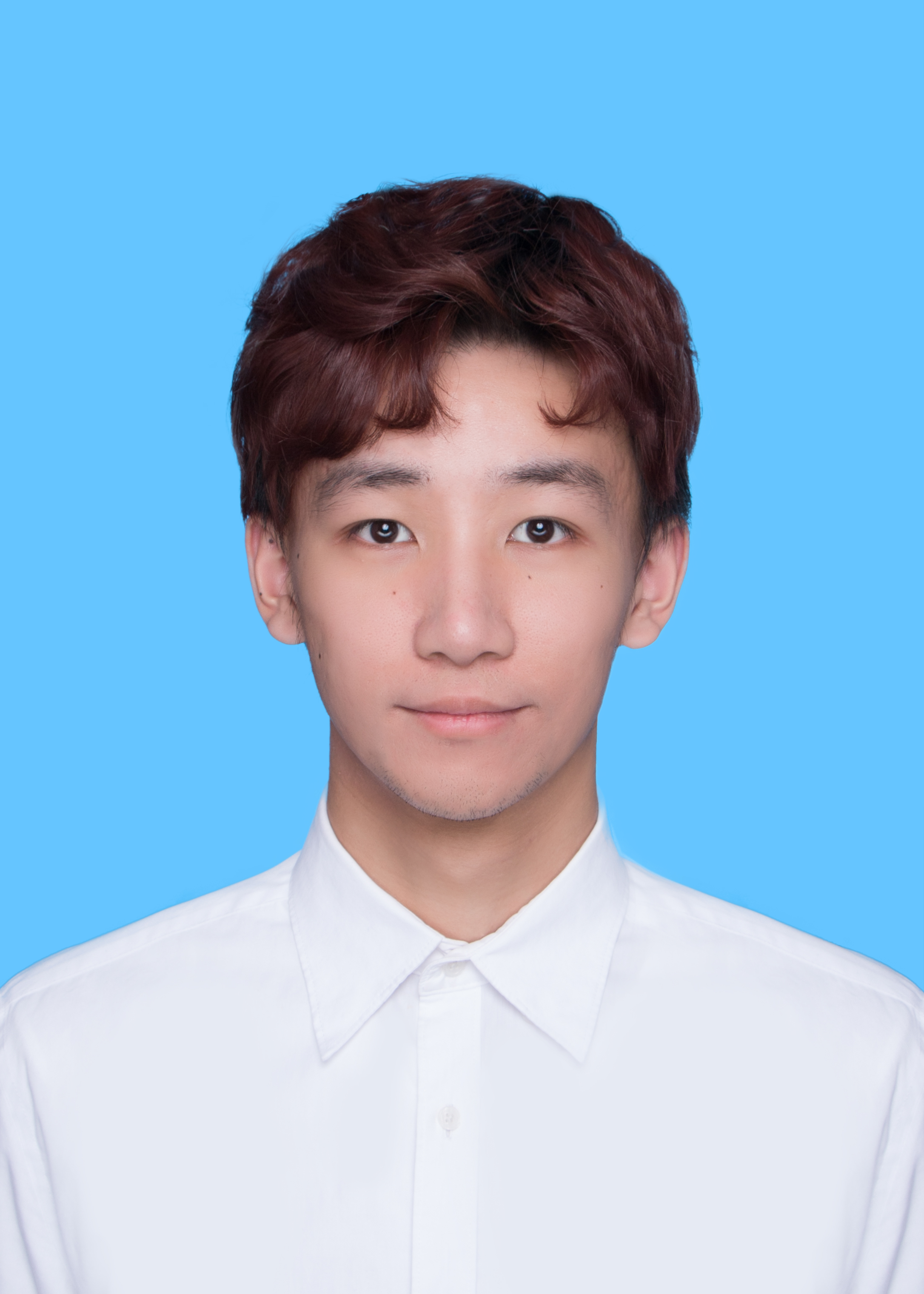}}]{Ziyan Zeng}
received the bachelor’s degree from the College of Artificial Intelligence, Southwest University, Chongqing, China. He is currently pursuing the master’s degree in computer science. His research interests include complex networks, evolutionary games, stochastic processes, and nonlinear science. 
\end{IEEEbiography}
\begin{IEEEbiography}[{\includegraphics[width=1in,height=1.25in,clip,keepaspectratio]{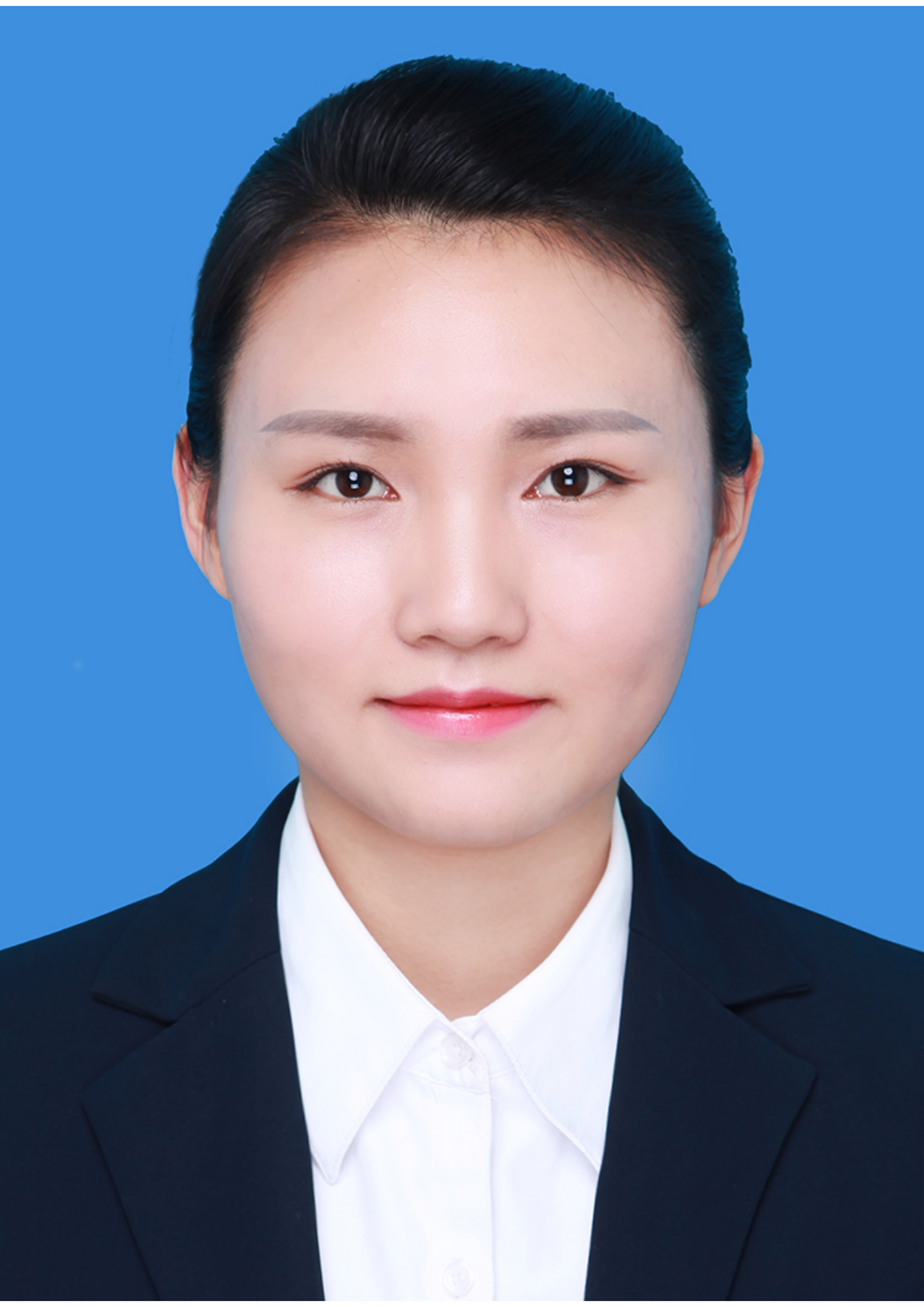}}]{Qin Li} 
received the M.S. degree in Public Administration from the University of Electronic Science and Technology of China in 2018. She is currently pursuing the Ph.D. degree with School of Public Policy and Administration, Chongqing University, China. Her research interests include social computing, complex networks, and evolutionary games.
\end{IEEEbiography}
\begin{IEEEbiography}
[{\includegraphics[width=1in,height=1.25in,clip,keepaspectratio]{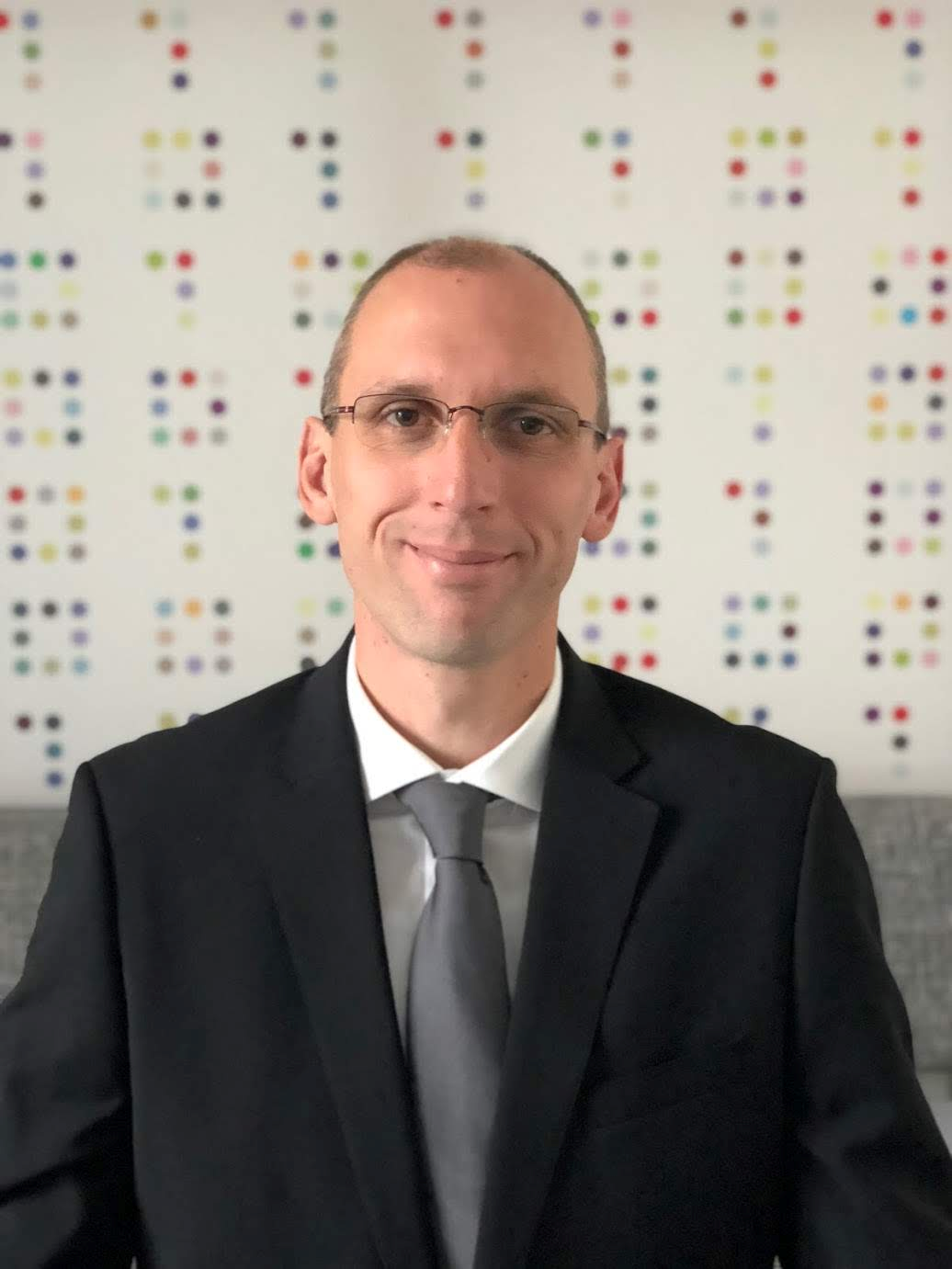}}]{Matja\v{z} Perc}
is professor of physics at the University of Maribor. He is a member of Academia Europaea and the European Academy of Sciences and Arts, and among top 1\% most cited physicists according to 2019, 2020, 2021, 2022 and 2023 Clarivate Analytics data. He is also the 2015 recipient of the Young Scientist Award for Socio and Econophysics from the German Physical Society, and the 2017 USERN Laureate. In 2018 he received the Zois Award, which is the highest national research award in Slovenia. In 2019 he became Fellow of the American Physical Society. Since 2021 he is Vice Dean of Natural Sciences at the European Academy of Sciences and Arts.
\end{IEEEbiography}
\begin{IEEEbiography}
[{\includegraphics[width=1in,height=1.25in,clip,keepaspectratio]{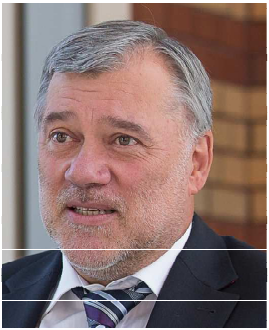}}]{J\"{u}rgen Kurths}
received the B.S. degree in mathematics from the University of Rostock, Rostock, Germany, the Ph.D. degree from the Academy of Sciences, German Democratic Republic, Berlin, Germany, in 1983, the Honorary degree from N.I. Lobachevsky State University, Nizhny Novgorod, Russia in 2008, and the Honorary degree from Saratow State University, Saratov, Russia, in 2012. 

From 1994 to 2008, he was a Full Professor with the University of Potsdam, Potsdam, Germany. Since 2008, he has been a Professor of Nonlinear Dynamics with Humboldt University, Berlin, and the Chair of the Research Domain Complexity Science, Potsdam Institute for Climate Impact Research, Potsdam. He has authored more than 700 papers, which are cited more than 53 000 times (H-index: 106). His main research interests include synchronization, complex networks, time-series analysis, and their applications. 

Dr. Kurths was the recipient of the Alexander von Humboldt Research Award from India, in 2005, and from Poland in 2021, the Richardson Medal of the European Geophysical Union in 2013, and the Eight Honorary Doctorates. He is a Highly Cited Researcher in Engineering. He is an Editor-in-Chief of \textit{CHAOS} and on the editorial boards of more than ten journals. He is a Fellow of the American Physical Society, the Royal Society of 1023 Edinburgh, and the Network Science Society. He is a member of the Academia 1024 Europaea.
\end{IEEEbiography}
\end{document}